\documentclass[aps,prd,superscriptaddress,nofootinbib,11pt]{revtex4}
\usepackage[english]{babel}
\usepackage[utf8]{inputenc}
\usepackage{graphicx}   
\usepackage{slashed}
\usepackage{epstopdf}
\usepackage{verbatim}   
\usepackage{color}      
\usepackage{subfigure}  
\usepackage{multirow}
\usepackage{hyperref}   
\usepackage{float}
\usepackage{epsfig,rotating}
\usepackage{amsmath,amssymb}
\usepackage{dsfont}
\restylefloat{table}
\numberwithin{equation}{section}

\newcommand{\vx}{\vec{x}}
\newcommand{\vy}{\vec{y}}
\newcommand{\vp}{\vec{p}}
\newcommand{\vq}{\vec{q}}
\newcommand{\vk}{\vec{k}}

\newcommand{\op}{\mathcal{O}_\chi}

\newcommand{\be}{\begin{equation}}
\newcommand{\ee}{\end{equation}}
\newcommand{\bea}{\begin{eqnarray}}
\newcommand{\eea}{\end{eqnarray}}

\newcommand{\ket}[1]{|#1\rangle}

\newcommand{\A}{\mathcal{A}}
\newcommand{\R}{\mathcal{R}}
\newcommand{\J}{\mathcal{J}}

\begin{document}
\title{Brownian Axion-like particles.}

\author{Shuyang Cao}
\email{shuyang.cao@pitt.edu} \affiliation{Department of Physics and
Astronomy, University of Pittsburgh, Pittsburgh, PA 15260}
\author{Daniel Boyanovsky}
\email{boyan@pitt.edu} \affiliation{Department of Physics and
Astronomy, University of Pittsburgh, Pittsburgh, PA 15260}

 \date{\today}

\begin{abstract}
We study the non-equilibrium dynamics of a pseudoscalar axion-like particle (ALP) weakly coupled to degrees of freedom in thermal equilibrium by obtaining its reduced density matrix. Its time evolution is determined by the in-in effective action which we obtain to leading order in the (ALP) coupling but to \emph{all orders} in the couplings of the bath  to other fields within or beyond the standard model. The effective  equation of motion for the (ALP)  is a Langevin equation with noise and friction kernels obeying the fluctuation dissipation relation. A ``misaligned'' initial condition yields  damped coherent oscillations, however, the (ALP)  population increases towards thermalization with the
bath. As a result, the     energy density features a mixture of a cold component from misalignment and a hot component  from thermalization with proportions that vary in time $(cold)\,e^{-\Gamma t}+(hot)\,(1-e^{-\Gamma t})$, providing a scenario wherein the ``warmth'' of the dark matter evolves in time from colder to hotter.  As a specific example we consider the (ALP)-photon  coupling $g a \vec{E}\cdot \vec{B}$ to lowest order, valid   from recombination onwards. For $T \gg m_a$ the long-wavelength relaxation rate is substantially enhanced $\Gamma_T  = \frac{g^2\,m^2_a\,T}{16\pi} $. The ultraviolet divergences of the (ALP) self-energy require higher order derivative terms in the effective action.  We find   that at high temperature, the finite temperature effective mass of the (ALP) is $m^2_a(T) = m^2_a(0)\Big[ 1-(T/T_c)^4\Big]$, with $T_c \propto \sqrt{m_a(0)/g}$, \emph{suggesting} the possibility of an inverted phase transition, which when combined with higher derivatives may possibly indicate  exotic new phases. We discuss possible cosmological consequences on structure formation, the effective number of relativistic species and birefringence of the cosmic microwave background.
\end{abstract}


\maketitle

\section{Introduction}
The axion,  introduced in Quantum Chromodynamics (QCD) as a solution of the strong CP problem\cite{PQ,weinaxion,wil} may be produced non-thermally in the Early Universe, for example by a misalignment mechanism and  is recognized as a potentially viable  cold dark matter candidate\cite{pres,abbott,dine}. Extensions beyond the standard model can accommodate pseudoscalar particles with properties similar to the QCD axion, namely axion-like-particles (ALP) which can also be suitable dark matter candidates\cite{banks,ringwald,marsh,sikivie1,sikivie2}, in particular as candidates for ultra light dark matter\cite{fuzzy,uldm}. Constraints on the mass and couplings of ultra light (ALP)\cite{marsh,sikivie1,sikivie2,banik} are being established  by various experiments\cite{cast,admx,graham}. There are two important features that characterize (ALP), i) a misalignment mechanism results in coherent oscillations of the expectation value of the (ALP) field which gives rise to the contribution to the energy density as a cold dark matter component\cite{pres,abbott,dine,marsh,sikivie1,sikivie2,turner}, ii) its pseudoscalar nature leads to an interaction between the (ALP) and photons  or  gluons via pseudoscalar composite operators of gauge fields, such as $\vec{E}\cdot\vec{B}$ in the case of the (ALP)-photon interaction and $G^{\mu\nu;b}\widetilde{G}_{\mu\nu;b}$ in the case of gluons, which allows an (ALP) to decay into two photons or gluons. The effect of this decay process in the evolution of (ALP) condensates has been studied in refs.\cite{sigl,arza,dashin} including stimulated decay in a photon background.

\vspace{1mm}

\textbf{Motivation and objectives:} In this article we study the non-equilibrium dynamics of coherent oscillations of (ALP) coupled to generic environmental fields in thermal equilibrium by obtaining the non-equilibrium in-in effective action from which we derive the effective equations of motion of (ALP) condensates.

A simple example highlights our main motivation and objectives: consider the textbook situation of a particle in an harmonic potential immersed in a heat bath in equilibrium. The interaction of the particle with the bath degrees of freedom induce two main modifications to the equations of motion of the particle: i) a friction term arising from energy transfer with the bath degrees of freedom, ii) a stochastic noise term arising from the random ``kicks'' that the environment gives the particle. This is the basis of Brownian motion and the effective equation  of motion of the Brownian particle is a \emph{Langevin} equation:

\be \ddot{x}(t) + \gamma \dot{x}(t)+ \omega^2 x(t) = \xi(t)\,,\label{lanclas} \ee with $\xi$ a stochastic noise with a (generally) Gaussian probability distribution function   yielding the (classical) averages and correlations
\be \langle \langle \xi(t) \rangle \rangle = 0~~;~~ \langle \langle \xi(t) \xi(t') \rangle \rangle = 2\gamma k_B\,T \delta(t-t')\,. \label{claslan}\ee The relation between the noise correlation function and the friction coefficient  in  (\ref{claslan}) is the (classical) fluctuation dissipation relation, a direct consequence of the bath degrees of freedom being in thermal equilibrium. As a result,  whereas the (stochastic) average $\langle \langle x(t) \rangle \rangle  \propto e^{-\gamma t/2} \cos[\sqrt{\omega^2-\frac{\gamma^2}{4}}\,t]$, the mean square fluctuation
$\langle \langle x^2(t) \rangle \rangle ~{}_{\overrightarrow{t\gg 1/\gamma}}~ k_B T/\omega^2 $, this is simply classical equipartition, namely the Brownian particle reaches  thermal equilibration with the bath on a relaxation time scale $\propto 1/\gamma$.

 This simple illustrative example motivates our study in this article, namely to understand the  {effective dynamics} of (ALP) when they are coupled to a bath of other degrees of freedom in (local) thermal equilibrium. The familiar example of a Brownian particle in a heat bath suggests that the \emph{effective} equations of motion of a coherent (ALP) condensate should be akin to a Langevin equation with a friction and noise term related by a fluctuation dissipation relation as a consequence of the bath degrees of freedom with which the (ALP) interacts being in thermal equilibrium. Our objective is precisely to derive, and solve such equation and explore its consequences by implementing the methods of non-equilibrium field theory. For this purpose, we adapt the seminal formulation of quantum Brownian motion\cite{feyver,leggett,ford,schmid} to the realm of non-equilibrium quantum field theory\cite{beilok,hupaz,calhu,das,boynoneq}. This is achieved in the  in-in or Schwinger-Keldysh\cite{schwinger,keldysh,maha,beilok} formulation of time evolution in quantum field theory. Unlike the in-out formulation, the in-in formulation yields causal, retarded equations of motion\cite{jordan,galley,boynoneq,boylee}.

 The objectives of this study are twofold: \textbf{i:)} to obtain the time evolution of a reduced density matrix,  non-equilibrium effective action, equations of motion and correlation functions for (ALP) particles   weakly coupled to degrees of freedom in thermal equilibrium. We first consider a generic model  with coupling of the form $\propto g a \mathcal{O}$ with $g$ a weak  coupling, $a$ the (ALP) field and $\mathcal{O}$ a composite pseudoscalar operator associated with the bath degrees of freedom.  We obtain the effective action to leading order in the coupling  ($\mathcal{O}(g^2)$), and \emph{to all orders} of the couplings of the environmental fields to other degrees of freedom within or beyond the standard model.      \textbf{ii:)} to apply the general results to the relevant case of (ALP)-photon interaction with $\mathcal{O} = \vec{E}\cdot \vec{B}$ where the radiation field in thermal equilibrium is identified with the cosmic microwave background (CMB).

 In this article we address these objectives in Minkowski space-time, obtaining  the (non-equilibrium) effective action and effective equations of motion for (ALP) to order $g^2$ in the  (weak) coupling $g$ and arbitrary (ALP) mass $m_a$ as a prelude to extending the methods  to an expanding cosmology and exploring phenomenological consequences and constraints in future work. Furthermore, in this study we do not adopt a particular set of parameters for (ALP) couplings and mass, nor on possible bounds on these. Our main focus is to study the general aspects of the effective dynamics resulting from these interactions under the sole assumption of weak coupling between the (ALP) and  degrees of freedom of the standard model and that the latter are in thermal equilibrium.

 \vspace{1mm}

 \textbf{Summary of results:} We study the time evolution of an initially prepared density matrix describing   a misaligned initial state for the (ALP) and an equilibrium thermal bath of generic fields coupled to the (ALP). Tracing over the bath fields yields a reduced density matrix for the (ALP) field whose time evolution is determined by the non-equilibrium effective action which is obtained up to $\mathcal{O}(g^2)$ but to all orders in the couplings of the bath degrees of freedom to other degrees of freedom within or beyond the standard model.

 The (ALP)  equations of motion obtained from the non-equilibrium effective action are stochastic of the Langevin type (\ref{lanclas}) with a friction kernel determined by the retarded (ALP) self energy (a manifestation of radiation reaction) and a Gaussian noise whose two point correlation function is related to the self-energy via a generalized fluctuation dissipation relation. This is   a consequence of the bath degrees of freedom being in thermal equilibrium. Hence, the notion of \emph{Brownian ALP's}. In ref.\cite{mottola} a local friction coefficient for the   equation of motion of the expectation value of the axion field was obtained as a consequence of sphaleronlike transitions in high temperature QCD. Thermal friction from the (ALP) coupling to high temperature plasmas has also been discussed in refs.\cite{friction,friction1}, and thermalization has been studied in refs.\cite{buch,masso}.  Our approach is very different in that we obtain the in-in non-equilibrium effective action which allows us to obtain the full equation of motion including the noise term and directly show that the self-energy contribution which yields the ``friction'' term and the noise correlation functions  are related by fluctuation dissipation.
 The noise term is of paramount importance in obtaining \emph{correlations} of the (ALP) field, and as a consequence of the noise term we find that the processes that lead to ``friction'' and  damping of the misaligned expectation value are the \emph{same} as those leading to thermalization with the environment  on similar time scales, thereby providing a direct bridge between damping of a coherent condensate and thermalization.  To the best of our knowledge, this approach, which explains both aspects of (ALP's) non-equilibrium dynamics, has not yet been implemented for (ALP).   A corollary of this important result, relevant for dark matter, is that
  the energy density features a mixture of a ``cold'' and ``hot'' components whose relative weight vary in time: $\mathcal{E} = (\mathrm{cold})\,e^{-\Gamma t} + (\mathrm{hot})\,(1-e^{-\Gamma t})$ where the ``cold'' component corresponds to the damped coherent oscillations arising from a misaligned initial condition and the hot component to the approach to thermalization and is a consequence of the stochastic noise. The relaxation rate $\Gamma$ is determined by the imaginary part of the (ALP) self-energy and is a result of (stimulated) emission and absorption processes with the bath.

 After obtaining the general results, we focus on the interaction of (ALP) with photons via the coupling $g \,a\, \vec{E}\cdot\vec{B}$. The coupling $g$ has dimensions of $1/(energy)$ resulting in a non-renormalizable interaction. Ultraviolet divergences necessitate the introduction of higher derivative terms of the (ALP) fields. Emission and absorption processes such as $a \leftrightarrow 2 \gamma$ yield a relaxation rate that is enhanced at high temperatures $T\gg m_a$ by a factor $ \propto T/m_a$. Furthermore, we find that the finite temperature contribution to the self-energy  yields a temperature dependent effective (ALP) mass $m^2_a(T) = m^2_a(0)\Big[ 1- (T/T_c)^4\Big]$ with $T_c \propto \sqrt{m_a(0)/g}$, this behavior of the effective mass \emph{suggests an inverted phase transition} and combined with the necessity for higher derivative terms in the effective Lagrangian points to the  possibility of novel phases and     Lifshitz   phase transitions\cite{lifshitz}.

In section (\ref{sec:noneLeff}) we obtain the   effective action out of equilibrium for (ALP) fields implementing the in-in Schwinger-Keldysh formulation of non-equiilibrium field theory for a generic interaction of the form $g a(x) \mathcal{O}(x)$, with  an initial density matrix for the (ALP) field that implements a misaligned initial condition and a thermal density matrix for the environmental fields. In this section we obtain general results: the Langevin equation of motion for the (ALP) field,  and the time dependent energy density with a cold component from misalignment and a hot component from thermalization.   In section (\ref{sec:EBint}) we focus on the interaction with photons, obtain the one loop self energy and noise correlation function at finite temperature, discuss renormalization issues and study their high and low temperature limits.

In section (\ref{sec:discussion}) we discuss the main aspects of the results and point out some caveats. In this section we argue that the results obtained   in the case of photon interactions are valid after recombination in the Early Universe and discuss possible cosmological consequences.  Our conclusions are summarized in section (\ref{sec:conclusions}). Several appendices include technical details.

\section{The   effective
action out of equilibrium}\label{sec:noneLeff}

 We study  the non equilibrium effective action of an axion-like field $ a(x)$ coupled to   generic fields $\chi(x)$ to which we refer as ``environmental'' fields via an operator $\op(x)$, with the Lagrangian density
  \be \mathcal{L}[a,\chi] = \frac{1}{2}\,\partial_\mu a(x) \partial^\mu a(x) - \frac{1}{2}\,m^2_{a}\,a^2(x) - g a(x)\,\op(x) + \mathcal{L}_{\chi} \label{lag}\ee where $\mathcal{L}_{\chi}$ is the Lagrangian density describing the ``environmental''  fields $\chi$, these fields could be the electromagnetic field, fermion or gluon fields. We  will first treat these fields generically to leading order in the coupling to exhibit the general form and properties of the effective action for (ALP) and then we will focus specifically on the case of the pseudoscalar coupling to the electromagnetic field, a hallmark of (ALP).

   The Lagrangian density (\ref{lag}) describes several relevant couplings of (ALP), such as
  \be \mathcal{L}_I = -g \,a(x)\,\vec{E}(x)\cdot \vec{B}(x)~;~     \mathcal{L}_I = -g_s \,a(x)\,G^{\mu \nu,b}(x)\widetilde{G}_{\mu \nu,b}(x)  ~;~\mathcal{L}_I = -g_{\psi} \,a(x) \overline{\Psi}(x)\gamma^5 \Psi(x) \cdots \,\label{inter} \ee where $\vec{E},\vec{B}$ are the electromagnetic fields, $G^{\mu \nu,b};\widetilde{G}^{\mu \nu,b}$ are the gluon field strength tensor and its dual respectively, and $\Psi(x)$ a fermionic field. Therefore the interaction in (\ref{lag}) describes a wide range of possible interactions of the (ALP) with other degrees of freedom which in this study are assumed to be in thermal equilibrium initially.

  Whereas the photon and gluon interactions are not renormalizable because the respective couplings $g,g_s$ feature dimensions $1/(energy)$, the coupling $g_{\psi}$ is dimensionless so the interaction with the fermionic pseudoscalar is renormalizable. This aspect will have important consequences as discussed below in section (\ref{sec:EBint}).

Upon evolving the total initial density matrix in time, the degrees of freedom $\chi$  with the generic operator $\op$ will be traced over to obtain a reduced density matrix for $a(x)$. We achieve this to leading order in the coupling $g$, but to all orders in the couplings of the $\chi$ fields with themselves or with other degrees of freedom within or beyond the standard model, except for the (ALP).

Although we are ultimately interested in obtaining an effective quantum field theory by tracing out  these degrees of freedom in an expanding  cosmology, in this study we focus on Minkowski space time as a first step towards extending these methods to cosmology.  We consider  the generic fields  $\chi$   as
a bath in thermal equilibrium.

The main strategy is to begin with an initial density matrix $\hat{\rho}(0)$ describing the (ALP) field and the environment, evolve it in time $\hat{\rho}(t) = U(t)\,\hat{\rho}(0)\,U^{-1}(t)$ with $U(t)$ the unitary time evolution operator for the (ALP)-environment, and trace over the environmental degrees of freedom yielding a reduced density matrix for the (ALP) fields, namely $\rho^{r}_a(t) = \mathrm{Tr}_{\chi}\hat{\rho}(t)$. This is the in-in  or Schwinger-Keldysh\cite{schwinger,keldysh,maha,beilok} formulation of non-equilibrium quantum field theory, the time evolution of the reduced density matrix is determined by a non-equilibrium effective action that includes the effects of the environment via a non-local term known as  the influence functional\cite{feyver} in the
theory of quantum brownian motion. This effective action yields \emph{causal} equations of motion\cite{jordan,galley}, which turn out to be \emph{stochastic}, akin to a Langevin equation with noise and dissipation terms that are related by a general fluctuation dissipation relation, a consequence of the environmental bath being in thermal equilibrium.

 The reduced density matrix can be
represented by a path integral in terms of the non-equilibrium
effective action that includes the influence functional.  This
method has been used previously to study quantum brownian
motion\cite{beilok,feyver,leggett,ford,schmid,hupaz,calhu,boynoneq} and for   studies of
quantum kinetics beyond the Boltzmann equation\cite{beilok,greiner,boykev}.

Let us consider the initial density matrix at a time
$t=0$ to be of the form
\begin{equation}
\hat{\rho}(0) = \hat{\rho}_{a}(0) \otimes
\hat{\rho}_{\chi}(0) \,.\label{inidensmtx}
\end{equation}

The initial density matrix $\hat{\rho}_a(0)$ is normalized so that  $\mathrm{Tr}_a \hat{\rho}_a(0) =1$ and that of the $\chi$ fields will be taken to
describe a statistical ensemble in thermal equilibrium at a temperature
$T=1/\beta$, namely

\begin{equation}\label{rhochi}
\hat{\rho}_{\chi}(0) = \frac{e^{-\beta\,H_{\chi}}}{\mathrm{Tr}_{\chi} e^{-\beta H_\chi}}\,,
\end{equation}

\noindent where $H_{\chi} $ is the total Hamiltonian for the
fields $\chi$, and may include other fields to which $\chi$ is coupled other than the (ALP), this possibility will be discussed further below.

The factorization of the initial density matrix is an assumption often explicitly or implicitly made in the literature, it can be relaxed by including initial correlations at the expense of daunting technical complications. We will not consider here this important case, relegating it to future study.

In the field basis the matrix elements of $\hat{\rho}_{a}(0)$ and $\hat{\rho}_{\chi}(0)$
are given by
\begin{equation}
\langle a |\hat{\rho}_{a}(0) | a'\rangle =
\rho_{a,0}(a ,a')~~;~~\langle \chi|\hat{\rho}_{\chi}(0) | \chi'\rangle =
\rho_{\chi,0}(\chi ;\chi')\,,
\end{equation} we emphasize that this is a \emph{functional} density matrix as the field has spatial arguments.
The density matrix for the (ALP) field $ a$   represents either  a pure state or more generally an initial
 statistical ensemble, whereas $\hat{\rho}_{\chi}(0)$ is given by eqn. (\ref{rhochi}).

  The physical situation described by (\ref{rhochi}) is that
  of a  field (or fields)  in thermal equilibrium at a temperature
  $T=1/\beta$, namely a heat bath,  which is put in contact with another system, here represented by the field $a$.
   Once the system  and bath are put in contact their mutual interaction will
  evolve the initial state out of equilibrium because the initial density matrix does not commute with the total Hamiltonian with interactions.

To obtain the effective action  out of equilibrium for the (ALP)  field $a$  we   evolve the initial density matrix in time and trace over the ``bath'' degrees of freedom, leading to a reduced density matrix for the   field $a$, from which
 we can compute its expectation values or correlation functions as a function of time.

The time evolution of the initial density matrix is given by

\begin{equation}
\hat{\rho}(t)= U(t)\hat{\rho}(0)U^{-1}(t)\,, \label{rhooft}
\end{equation}
where
\begin{equation}
U(t) = e^{-iHt}\,. \label{unitimeop} \ee
The total Hamiltonian $H$ is given by
\begin{equation}
H=H_{0 a} + H_{\chi}+g\int d^3x  ~ a(x)~\op(x)\,, \label{hami}
\end{equation}
and $H_{0a},H_{\chi}$  are the   Hamiltonians for   the respective fields.

The \emph{reduced} density matrix for the (ALP) field is obtained by tracing over the $\chi$ degrees of freedom as
\be \rho^{r}_a(t) = \mathrm{Tr}_{\chi} U(t) \hat{\rho}(0)\,U^{-1}(t)\,.  \label{rored}\ee To extract the non-equilibrium effective action for the (ALP) it is more convenient to obtain the density matrix elements in field space, namely

\be    \rho(a_f,\chi_f;a'_f,\chi'_f;t)   =        \langle a_f;\chi_f|U(t)\hat{\rho}(0)U^{-1}(t)|a'_f;\chi'_f\rangle \,,\label{evolrho} \ee from which the reduced density matrix elements are
\be \rho^r(a_f ;a'_f,;t) = \int D\chi_f \,\langle a_f;\chi_f|U(t)\hat{\rho}(0)U^{-1}(t)|a'_f;\chi_f\rangle \,. \label{redmtxel}\ee

With the functional integral representation
\bea \langle a_f;\chi_f|U(t)\hat{\rho}(0)U^{-1}(t)|a'_f;\chi'_f\rangle  & = &  \int Da_i D\chi_i Da'_i D\chi'_i ~ \langle a_f;\chi_f|U(t)|a_i;\chi_i\rangle\,\rho_{a,0}(a_i;a'_i)\, \otimes   \nonumber \\ & & \rho_{\chi,0}(\chi_i;\chi'_i) \,
 \langle a'_i;\chi'_i|U^{-1}(t)|a'_f;\chi'_f\rangle \,,\label{evolrhot}\eea it follows that the reduced
 density matrix elements are
\bea \rho^r(a_f ;a'_f,;t) & = & \int D\chi_f   \int Da_i D\chi_i Da'_i D\chi'_i ~ \langle a_f;\chi_f|U(t)|a_i;\chi_i\rangle\,\rho_{a,0}(a_i;a'_i)\, \otimes   \nonumber \\ & & \rho_{\chi,0}(\chi_i;\chi'_i) \,
 \langle a'_i;\chi'_i|U^{-1}(t)|a'_f;\chi_f\rangle \,.\label{redfun} \eea

  The $\int Da$ etc, are functional integrals where the spatial argument has been suppressed. The matrix elements of the time evolution forward and backward can be written as path integrals, namely
 \bea   \langle a_f;\chi_f|U(t)|a_i;\chi_i\rangle  & = &    \int \mathcal{D}a^+ \mathcal{D}\chi^+\, e^{i \int d^4 x \mathcal{L}[a^+,\chi^+]}\label{piforward}\\
 \langle a'_i;\chi'_i|U^{-1}(t)|a'_f;\chi'_f\rangle &  =  &   \int \mathcal{D}a^- \mathcal{D}\chi^-\, e^{-i \int d^3 x \mathcal{L}[a^-,\chi^-]}\label{piback}
 \eea where we use the shorthand notation
 \be \int d^4 x \equiv \int_0^t dt \int d^3 x \,,\label{d4xdef}\ee
 $ \mathcal{L}[a,\chi] $ is given by (\ref{lag})   and
 the boundary conditions on the path integrals are
  \bea     a^+(\vec{x},t=0)=a_i(\vec{x})~;~
 a^+(\vec{x},t)  &  =  &   a_f(\vec{x})\,,\label{piforwardbc}\\   \chi^+(\vec{x},t=0)=\chi_i(\vec{x})~;~
 \chi^+(\vec{x},t) & = & \chi_f(\vec{x}) \,,\label{chipfbc}  \\
     a^-(\vec{x},t=0)=a'_i(\vec{x})~;~
 a^-(\vec{x},t) &  = &    a'_f(\vec{x})\,,\label{aminbc} \\   \chi^-(\vec{x},t=0)=\chi'_i(\vec{x})~;~
 \chi^-(\vec{x},t) & = & \chi'_f(\vec{x}) \,.\label{pibackbc}
 \eea

The field variables $a^\pm, \chi\pm$ along the forward ($+$) and backward ($-$) evolution branches are recognized as those necessary for the in-in or  Schwinger-Keldysh\cite{schwinger,keldysh,maha,beilok} closed time path approach to the time evolution of a density matrix.

 The reduced density matrix for the light field $a$ (\ref{redfun}),    can be written as
 \be   \rho^{r}(a_f,a'_f;t) = \int Da_i   Da'_i  \,  \mathcal{T}[a_f,a'_f;a_i,a'_i;t] \,\rho_a(a_i,a'_i;0)\,, \ee
where the time evolution kernel is given by
\be \mathcal{T}[a_f,a_i;a'_f,a'_i;t] = {\int} \mathcal{D}a^+ \, \int \mathcal{D}a^- \, e^{i  \int d^4x \left[\mathcal{L}_0[a^+]-\mathcal{L}_0[a^-]\right]}\,e^{i\mathcal{I}[a^+;a^-]}\,, \ee
from which the \emph{in-in effective action} out of equilibrium is identified as
\begin{equation}
    {S}_{eff}[a^+,a^-] = \int^t_0 dt \int d^3 x \Big\{ \mathcal{L}_0[a^+]-\mathcal{L}_0[a^-] +\mathcal{I}[ a^+, a^-] \Big\} \,,\label{Leff}
\end{equation} where
$\mathcal{I}[a^+;a^-]$ is the \emph{influence action}\cite{feyver} obtained by tracing over the $\chi$ degrees of freedom,
\be
    e^{i\mathcal{I}[a^+;a^-]} = \int D\chi_i \,D\chi'_i D\chi_f  \int \mathcal{D}\chi^+ \int \mathcal{D}\chi^- \, e^{i  \int d^4x \left[\mathcal{L}[\chi^+]-g a^+\,\op^+\right]}~   e^{-i  \int d^4x \left[\mathcal{L}[\chi^-]-g  a^- \,\op^-\right]}\,\rho_{\chi}(\chi_i,\chi'_i;0)
    \label{inffunc}
\ee

The path integral representations for both $\mathcal{T}[a_f,a_i;a'_f,a'_i;t]$ and $\mathcal{I}[a^+;a^-]$ feature the boundary conditions in (\ref{piforwardbc}-\ref{pibackbc}) except that we now set $\chi^\pm(\vec{x},t) = \chi_f(\vec{x})$ to trace over $\chi$ field.


In the above path integral defining the influence action  eqn. (\ref{inffunc}),  the (ALP) fields $ a^\pm(x) $ act as   \emph{external sources} (c-number) coupled to the   operator $\mathcal{O}_\chi$. Therefore, it is straightforward to conclude that the right hand side of eqn. (\ref{inffunc}) is the path integral representation of the trace over the environmental fields coupled to external sources $a^\pm$, namely
\be e^{i\mathcal{I}[a^+;a^-]} = \mathrm{Tr}_{\chi} \Big[ \mathcal{U}(t;a^+)\,\rho_\chi(0)\,  \mathcal{U}^{-1}(t;a^-) \Big]\,, \label{trasa}\ee where   $\mathcal{U}(t;a^\pm)$ is the   time evolution operator in the $\chi$ sector in presence of \emph{external sources} $a^\pm$, i.e.  \be \mathcal{U}(t;a^+) = T\Big( e^{-i \int_0^t H_\chi[a^+(t')]dt'}\Big) ~~;~~
\mathcal{U}^{-1}(t;a^-) = \widetilde{T}\Big( e^{i \int_0^t H_\chi[a^-(t')]dt'}\Big)\,,\label{us} \ee
with \be H_\chi[a^\pm(t)] = H_{\chi}+g\,\int d^3x \, a^\pm(\vec{x},t)\op(\vx)  \label{totiH} \ee   and $\widetilde{T}$ is the \emph{anti-time evolution operator} describing   evolution backward in time, it is defined by $\widetilde{T}(A(t_1)B(t_2)) = A(t_1) B(t_2)\Theta(t_2-t_1)+B(t_2)A(t_1)\Theta(t_1-t_2)$.

The calculation of the influence action is facilitated by passing to the interaction picture for the Hamiltonian $H_\chi[a^\pm(t)]$, defining
\be  \mathcal{U}(t;a^\pm) = e^{-i H_{\chi}\,t} ~ \mathcal{U}_{ip}(t;a^\pm) \label{ipicture} \ee and the $e^{\pm i H_{\chi}\,t}$ cancel out in the trace in (\ref{trasa}), since $\mathcal{U}(t;a^\pm)$ is the time evolution operator in presence of \emph{external} sources $a^{\pm}(\vx,t)$ for the $\chi$ sector, it follows that
 \bea   \mathcal{U}_{ip}(t;a^+)  & = &  1-i\,g \,  \int d^4 x' a^+(\vx,t')\op(\vx,t') \nonumber \\ & - & \frac{g^2}{2}\,  \int d^4 x_1\,  \int d^4 x_2 T \Big(a^+(\vx_1,t_1)\op(\vx_1,t_1)a^+(\vx_2,t_2)\op(\vx_2,t_2)\Big)+\cdots \label{uplus}\eea
 \bea  \mathcal{U}^{-1}_{ip}(t;a^-)  & =  &  1+i\,g\,  \int d^4 x' a^-(\vx',t')\op(\vx',t') \nonumber \\ &  - & \frac{g^2}{2}\,  \int d^4 x_1\, \int d^4 x_2 \widetilde{T}\Big(a^-(\vx_1,t_1)\op(\vx_1,t_1)a^-(\vx_2,t_2)\op(\vx_2,t_2)\Big)+\cdots \,,\label{umin} \eea where $\op(\vx,t)$ is in the Heisenberg picture of $H_{\chi}$.

Now the trace (\ref{trasa}) can be obtained systematically in perturbation theory in $g$ from which we obtain the influence functional. Up to $\mathcal{O}(g^2)$  we find
\bea \mathcal{I}[a^+,a^-] & = &    - g \int d^4x \Big( a^+(x)-a^-(x)\Big)\,\langle \op(x)\rangle_\chi \nonumber + \\ & & \frac{i g^2 }{2} \int d^4x_1 \int d^4x_2 \Bigg\{ a^+(x_1) \,a^+(x_2)\,G_c^{++}(x_1-x_2)+ a^-(x_1)\,a^-(x_2)\,G_c^{--}(x_1-x_2) \nonumber \\
 & - & a^+(x_1)\,a^-(x_2)\,G_c^{+-}(x_1-x_2)- a^-(x_1)\,a^+(x_2)\,G_c^{-+}(x_1-x_2)\Bigg\}\,, \label{finF}\eea which is confirmed by expanding the left hand side  of (\ref{trasa}) and comparing to the right hand side. In this expression  the \emph{connected} correlation functions in the initial density matrix of the $\chi$ fields, namely $\rho_\chi(0)$ are given by
\begin{eqnarray}
&& G_c^{-+}(x_1-x_2) =   \langle
{\cal O}_{\chi}(x_1) {\cal O}_{\chi}(x_2)\rangle_\chi - \langle \mathcal{O}_{\chi}(x_1)\rangle_\chi  \langle \mathcal{O}_{\chi}(x_2)\rangle_\chi =   {G}_c^>(x_1-x_2) \,,\label{ggreat} \\&&  G_c^{+-}(x_1-x_2) =   \langle
{\cal O}_{\chi}(x_2) {\cal O}_{\chi}(x_1)\rangle_\chi - \langle \mathcal{O}_{\chi}(x_2)\rangle_\chi  \langle \mathcal{O}_{\chi}(x_1)\rangle_\chi =   {G}_c^<(x_1-x_2)\,,\label{lesser} \\&& G_c^{++}(x_1-x_2)
  =
{ G}_c^>(x_1-x_2)\Theta(t_1-t_2)+ {  G}_c^<(x_1-x_2)\Theta(t_2-t_1) \,,\label{timeordered} \\&& G_c^{--}(x_1-x_2)
  =
{ G}_c^>(x_1-x_2)\Theta(t_2-t_1)+ {  G}_c^<(x_1-x_2)\Theta(t_1-t_2)\,,\label{antitimeordered}
\end{eqnarray} in terms of fields in the Heisenberg picture of $H_\chi$, where
\be \langle (\cdots) \rangle = \mathrm{Tr}_\chi(\cdots)\rho_\chi(0)\,. \label{expec}\ee
Furthermore, for the case of hermitian operators $\mathcal{O}_\chi$ as considered here it follows that
\be G_c^>(x_1-x_2) = G_c^<(x_2-x_1)\,. \label{idengs}\ee

We highlight  that the correlation functions $G^{>},G^{<}$ are \emph{exact}, namely to \emph{all orders} in the couplings of the environmental fields $\chi$ that enter in $\mathcal{O}$ to all other fields to which it couples but the (ALP).

In the cases under consideration, we assume that the initial density matrix for the bath,    $\rho_\chi(0)=e^{-\beta H_{\chi}}$   is $CP$ invariant, for example in Quantum Electrodynamics where $\rho_{\chi}(0)$ describes blackbody radiation, for which  $\langle \vec{E}\cdot \vec{B}\rangle =0$.    Therefore   $\langle \mathcal{O}(x_{1,2})\rangle\equiv   0$ in the connected correlation functions (\ref{ggreat}-\ref{antitimeordered}), hence in what follows we suppress the subscript ``c'' in the correlation functions.

The influence action (\ref{finF}) becomes simpler by writing it  solely in terms of the two correlation functions $G^\lessgtr$, this is achieved by implementing the following steps:

\begin{itemize}
\item{In the term with $a^+(x_1)a^+(x_2)$: in the contribution $G^<(x_1-x_2)\Theta(t_2-t_1)$ (see eqn. (\ref{timeordered})) relabel $x_1 \leftrightarrow x_2$ and use the property (\ref{idengs}).  }
\item{ In the term with $a^-(x_1)a^-(x_2)$: in the contribution $G^>(x_1-x_2)\Theta(t_2-t_1)$ (see eqn. (\ref{antitimeordered})) relabel $x_1 \leftrightarrow x_2$ and use the property (\ref{idengs}). }
    \item{ In the term with $a^+(x_1)a^-(x_2)$: multiply $G^<(x_1-x_2)$ by $\Theta(t_1-t_2)+\Theta(t_2-t_1)=1$ and in the term with $\Theta(t_2-t_1)$ relabel $ x_1 \leftrightarrow x_2$ and use the property (\ref{idengs}). }
 \item{ In the term with $a^-(x_1)a^+(x_2)$: multiply $G^>(x_1-x_2)$ by $\Theta(t_1-t_2)+\Theta(t_2-t_1)=1$ and in the term with $\Theta(t_2-t_1)$ relabel $x_1 \leftrightarrow x_2$ and use the property (\ref{idengs}). }
\end{itemize}
We find
\bea \mathcal{I}[a^+, a^-]  & & =    i\,g^2\int d^4x_1 d^4x_2  \,\Bigg\{ a^+(\vx_1,t_1)a^+(\vx_2,t_2)\,G^>(x_1-x_2) +   a^-(\vx_1,t_1)a^-(\vx_2,t_2)\,G^<(x_1-x_2) \nonumber\\
  &- & a^+(\vx_1,t_1)a^-(\vx_2,t_2)\,G^<(x_1-x_2)  -   a^-(\vx_1,t_1)a^+(\vx_2,t_2)\,G^>(x_1-x_2)\Bigg\}\Theta(t_1-t_2) \label{Funravel}\eea where $G^{\lessgtr}$ are given by eqns. (\ref{ggreat},\ref{lesser}).  This is the general form of the influence function up to second order in the (ALP)-environment coupling but \emph{to all orders} in the couplings of the environmental fields that enter the composite operator $\mathcal{O}$ to \emph{any} other field. Notice that $\mathcal{I}[a^+, a^-]\Big|_{a^+=a^-}=0$ consistently with its definition given by eqn. (\ref{trasa}).  A graphical depiction of the influence action $\mathcal{I}[a^+,a^-]$ is displayed in fig.(\ref{fig:effaction}).

     \begin{figure}[h!]
\includegraphics[height=2.5in,width=2.5in,keepaspectratio=true]{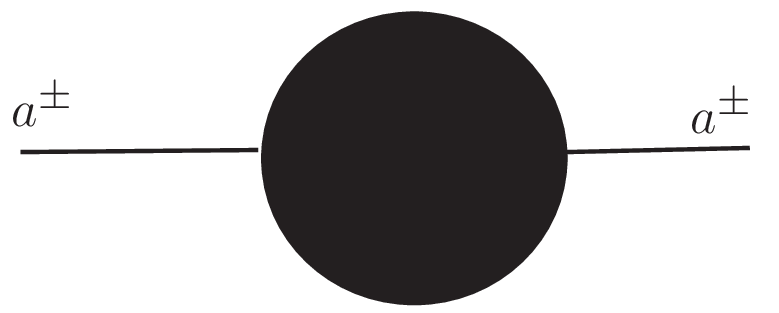}
\caption{A graphical depiction of $\mathcal{I}[a^+, a^-]$. The black circle denotes the correlation functions $G^{\lessgtr}$ to all orders in the couplings to degrees of freedom other than the (ALP).}
\label{fig:effaction}
\end{figure}

For example, for the (ALP)-photon interaction in eqn. (\ref{inter}), some of the correlations included in the influence action are displayed in fig. (\ref{fig:moreloops}): the one-loop diagram features free photon propagators, the two-loop diagram features a polarization correction to one of the propagators with electron-positron pairs in the thermal bath, this two- loops diagram features an extra power of $\alpha$ the fine structure constant. Similar diagrams with quarks and gluon loops are included for the   (ALP)-gluon interaction in (\ref{inter}). The black ``bubble'' symbolizes the $\langle \mathcal{O} \mathcal{O}\rangle$ correlation functions of the bath degrees of freedom in thermal equilibrium to all orders in their interactions, $\alpha, \alpha_s$ etc.

     \begin{figure}[h!]
\includegraphics[height=4in,width=4in,keepaspectratio=true]{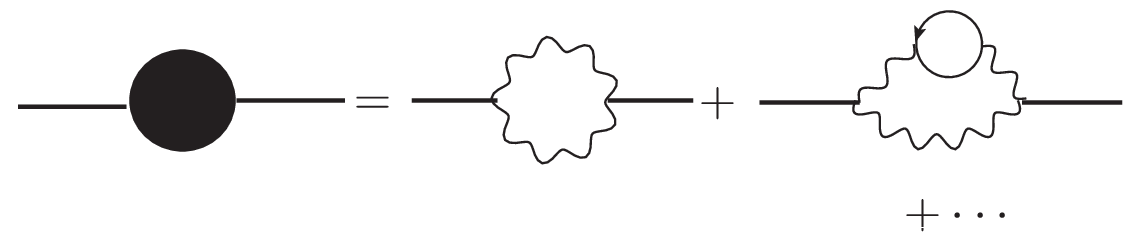}
\caption{  $\mathcal{I}[a^+, a^-]$ for the (ALP)-photon interaction in (\ref{inter}). The one loop diagram features free photon propagators, the two loop diagram includes the polarization from $e^+e^-$ pairs in the thermal plasma, etc.}
\label{fig:moreloops}
\end{figure}

  We can obtain expectation values and correlation functions of the (ALP) fields by including sources $J^{\pm}(x)$ with $\mathcal{L}_0(a^\pm)\rightarrow \mathcal{L}_0(a^\pm)+J^{\pm}(x)a^\pm(x)$ and defining the generating functional
  \be \mathcal{Z}[J^+,J^-] = \mathrm{Tr}\,\rho^r(J^+,J^-;t) =  \int Da_f Da_i   Da'_i  \,  {\int} \mathcal{D}a^+ \, \int \mathcal{D}a^- \, e^{i S_{eff}[a^+,J^+;a^-,J^-;t]} \,\rho_a(a_i,a'_i;0) \label{Zofj}\ee with the boundary conditions
  \bea &  &   a^+(\vec{x},t=0)=a_i(\vec{x})~;~
 a^+(\vec{x},t)  =   a_f(\vec{x}) \nonumber \\
&  &   a^-(\vec{x},t=0)=a'_i(\vec{x})~;~
 a^-(\vec{x},t)  =   a_f(\vec{x}) \,.\label{bctraza}\eea Expectation values or correlation functions of $a^{\pm}$ in the reduced density matrix are obtained as usual with variational derivatives with respect to the sources $J^\pm$.

\subsection{Effective equations of motion: Langevin equation}\label{subsec:langevin}
The effective action (\ref{Leff}) may be written in a manner more suitable to exhibit the equations of motion by introducing the Keldysh\cite{keldysh} variables
\be \mathcal{A}(\vx,t)= \frac{1}{2}\big( a^+(\vx,t)+a^-(\vx,t)\big)~~;~~\mathcal{R}(\vx,t)= \big( a^+(\vx,t)-a^-(\vx,t)\big)\label{kelvars} \,. \ee

The boundary conditions on the $a^\pm$ path integrals given by
(\ref{bctraza}) translate into the following boundary conditions on
the center of mass and relative variables
\begin{align}
    \A(\vec x,t=0)= \A_i \; \; &; \; \; \R(\vec x,t=0)=\R_i \,, \label{bcwig} \\
    \A(\vec{x},t=t_f) = a_f(\vec{x}) \; \; &; \; \; \R(\vec x,t=t_f )=0 \,. \label{Rfin}
\end{align}

 In terms of the center of mass and relative field variables, the effective action (\ref{Leff})  with the influence functional (\ref{finF}) becomes with $\omega^2_k = m^2_a + k^2$
\begin{align}
    iS_{eff}[\A,\R] =
    & - i \int d^3x \R_i(x) \dot{\A}(x,t=0) \nonumber \\
    & + i \int_0^t dt\, \sum_{\vec{k}} \left\{ - \R_{-\vec{k}} \left( \ddot{\A}_{\vec{k}}(t) + \omega_k^2 \A_{\vec{k}}(t) \right) + \A_{\vec{k}} \J_{-\vec{k}} \right\} \nonumber \\
    & - \int_0^t dt_1 \int_0^t dt_2 \left\{ \frac{1}{2} \R_{-\vec{k}}(t_1) \mathcal{N}_{\vec{k}}(t_1-t_2) \R_{\vec{k}}(t_2) + \R_{-\vec{k}} i \Sigma_{\vec{k}}^R(t_1 - t_2) \A_{\vec{k}}(t_2) \right\}
    \label{efflanwig}
\end{align} where we have integrated by parts and defined $ \J(x)= (J^+(x)-J^-(x))$, keeping solely the source conjugate to $\A$ because we are interested in expectation values and correlation functions of this variable only as discussed in detail below.

The kernels in the above effective Lagrangian are given by (see
eqns. (\ref{ggreat}-\ref{antitimeordered}))
\begin{eqnarray}
\mathcal{N}_k(t-t') & = &
\frac{g^2}{2} \left[   G^>(k;t-t')+{ G}^<(k;t-t') \right] \label{kernelkappa} \\
i\Sigma^{R}_k(t-t') & = &  g^2 \left[{ G}^>(k;t-t')-{
G}^<(k;t-t') \right]\Theta(t-t') \equiv
i\Sigma_{k}(t-t')\Theta(t-t') \label{kernelsigma}
\end{eqnarray} where $G^{<,>}(k;t-t')$ are the spatial Fourier transforms of the correlation functions in (\ref{ggreat}-\ref{antitimeordered}). In the exponential of the effective action $e^{iS_{eff}}$, the quadratic term in the relative variable $\R$ can be written as a functional integral over a noise variable $\xi$ as follows,

\begin{align}
    &\exp\left\{ -  \frac{1}{2}\,\int dt_1 \int dt_2   \,\R_{-\vec{k}}(t_1) \mathcal{N}_{\vec{k}}(t_1 - t_2) \R_{\vec{k}}(t_2) \right\}
    \nonumber \\
    & \qquad\qquad\qquad = \widetilde{C} \int \mathcal{D}\xi \, \exp\left\{ - \frac{1}{2} \int dt_1 \int dt_2 \, \xi_{-\vec{k}}(t_1) \mathcal{N}^{-1}_{\vec{k}}(t_1-t_2) \xi_{\vec{k}}(t_2) + i \int dt \xi_{-\vec{k}}(t) \R_{\vec{k}}(t) \right\}
    \label{nois}
\end{align}
where $\widetilde{C}$ is a normalization factor.

For the initial density matrix $\rho_a(a_i,a'_i;0)$ in (\ref{Zofj}) it proves convenient to write it in terms of the initial center of mass and relative variables $\A_i,\R_i$ as
\be \rho_a(a_i,a'_i;0) \equiv \rho_a(\A_i+\frac{\R_i}{2},\A_i-\frac{\R_i}{2};0) \label{rhoavars}\ee and introduce the functional Wigner transform\cite{zubairy}
\be W[\A_i,\pi_i] = \int D\R_i \, e^{-i \int d^3 x \pi_i(\vx) \,\R_i(\vx)}\,\rho_a(\A_i+\frac{\R_i}{2},\A_i-\frac{\R_i}{2};0) \,,\label{wigner}\ee which allows us to write (up to a normalization factor)
\be \rho_a(\A_i+\frac{\R_i}{2},\A_i-\frac{\R_i}{2};0)    = \int D\pi_i \, e^{i \int d^3 x  \pi_i(\vx) \,\R_i(\vx)}\, W[\A_i,\pi_i]
\,. \label{invwig}\ee

As it will become clear below, the Wigner transform naturally leads to an initial value problem wherein the evolution of the field is determined from initial conditions on its value and first time derivative.

Gathering these results together, we now write the generating functional (\ref{Zofj}) in terms of the Keldysh variables (\ref{kelvars}), with the effective action in these variables given by eqn. (\ref{efflanwig}), implementing the Wigner transform (\ref{invwig}) and using the representation (\ref{nois})
\begin{align}
    \mathcal{Z}[\mathcal{J}] =
    & \int D\A_f \int D\R_i\, D\A_i\, D\pi_i \int D\A\, D\R\, D\xi\; W[\A_i,\pi_i] \times P[\xi] \times \exp\left\{ i \int dt \sum_{\vec{k}} \A_{\vec{k}}(t) \J_{-\vec{k}}(t) \right\}
    \nonumber \\
    & \times \exp\left\{ -i \int dt \sum_{\vec{k}} \left[ \R_{-\vec{k}}(t) \left( \ddot{\A}_{\vec{k}}(t) + \omega_{\vec{k}}^2 \A_{\vec{k}}(t) + \int_0^t \Sigma_{\vec{k}}(t-t') \A_{\vec{k}}(t') d't - \xi_{\vec{k}}(t) \right) \right] \right\}
    \nonumber \\
    & \times \exp\left\{ i \sum_{\vec{k}} \R_i(-\vec{k}) \left( \pi(\vec{k}) - \dot{\A}_i(\vec{k}) \right)  \right\}
    \label{Zetafinal}
\end{align} where the noise probability distribution function
\begin{equation}
    P[\xi] = \widetilde{N}\, \prod_{\vec{k}} \exp\left\{ -\frac{1}{2} \int  dt_1 \int  dt_2\, \xi_{-\vec k}(t_1)\,{\mathcal{N} }^{-1}_k(t_1-t_2)\,\xi_{\vec k}(t_2) \right\}\,.
    \label{noispdf}
\end{equation}
The generating functional $\mathcal{Z}[\J]$ is the final form of the time evolved reduced density matrix   after tracing over the bath degrees of freedom. Variational derivatives with respect to the source $\J$ yield  the correlation functions of the Keldysh center of mass variable $\A$.

Carrying out the functional integrals over $\R_i(\vec{k})$ and $\R_{\vec{k}}(t)$ yields a more clear form, namely
\begin{align}
    \mathcal{Z}[\mathcal{J}] \propto
    & \int D\A_f \int D\A_i\, D\pi_i \int D\A\, D\xi\; W[\A_i,\pi_i] \times P[\xi] \times \exp\left\{ i \int dt \sum_{\vec{k}} \A_{\vec{k}}(t) \J_{-\vec{k}}(t) \right\}
    \nonumber \\
    & \times \prod_{\vec{k}} \delta\left[ \ddot{\A}_{\vec{k}}(t) + \omega_{\vec{k}}^2 \A_{\vec{k}}(t) + \int_0^t \Sigma_{\vec{k}}(t-t') \A_{\vec{k}}(t') d't - \xi_{\vec{k}}(t) \right]
      \times \prod_{\vec{k}} \delta\left[ \pi(\vec{k}) - \dot{\A}_i(\vec{k}) \right]\,.
    \label{Zetadelta}
\end{align}
The functional delta functions clearly determine   the field configurations that contribute  to the generating functional $\mathcal{Z}[\mathcal{J}]$:
\begin{itemize}

 \item The equation of motion of $\A_{\vec{k}}(t)$ is a   \emph{stochastic} Langevin equation, namely
    \begin{equation}
        \ddot{\A}_{\vec{k}}(t) + \omega_{\vec{k}}^2 \A_{\vec{k}}(t) + \int_0^t \Sigma_{\vec{k}}(t-t') \A_{\vec{k}}(t') d't = \xi_{\vec{k}}(t)\,.\label{langevin}
    \end{equation}
    Note that this equation of motion involves the \emph{retarded} self-energy, thereby defining a causal initial value problem, this is a distinct consequence of the in-in formulation of time evolution.

    \item The initial conditions of $\A_{\vec{k}}$ satisfy
    \begin{equation}
        \A_{\vec{k}}(t=0) = \A_{i,\vec{k}} \qquad;\qquad \dot{\A}_{\vec{k}}(t=0) = \pi_{i,\vec{k}}\,, \label{inicons}
    \end{equation}
    where $\A_{i,\vec{k}},\pi_{i,\vec{k}}$ are drawn from the distribution function $W[\A_i,\pi_i]$ (i.e.,  the initial density matrix). This is one of the manifestations of stochasticity, and we use $\overline{(\cdots)}$ to denote averaging over the initial conditions (\ref{inicons})  with the distribution function $W[\A_i,\pi_i]$.

    \item The expectation value and correlations of the stochastic noise $\xi_{\vec{k}}(t)$ are determined by a Gaussian probability distribution $P[\xi]$, yielding
    \be
        \langle\langle \xi(\vx,t)  \rangle \rangle =0
        ~~;~~
        \langle\langle \xi_{\vk}(t) \xi_{\vk'}(t')  \rangle \rangle = \mathcal{N}_k(t-t')\,\delta_{\vk,-\vk'} \,, \label{noiscors}
    \ee
    where $\langle\langle \cdots \rangle\rangle$ means averaging weighted by $P[\xi]$. Since $P[\xi]$ is Gaussian, higher order correlation functions are obtained by implementing Wick's theorem. This averaging is the second manifestation of stochasticity.
\end{itemize}

Therefore, averaging over both the initial conditions with the Wigner distribution function, and the noise with $P[\xi]$,    is now denoted by $\overline{\langle\langle \big( \cdots \big) \rangle\rangle}$ and $\big( \cdots \big)$ is any functional of the initial conditions (\ref{inicons}) and $\xi$. These stochastic averages  yield  the expectation values and correlation functions of functionals of $\mathcal{A}$ obtained from variational derivatives with respect to $\mathcal{J}$.

It remains to relate observables to correlation functions of the Keldysh center of mass variable $\A$. The path integral representations for the forward and backward time evolution operators (\ref{evolrhot}, \ref{piforward},\ref{piback}) show that $a^+$ is associated with $U(t)$ and $a^-$ with $U^{-1}(t)$, hence it follows that  inside the path integral operators in the forward, backward and mixed forward-backward branches,

 \be A^+ B^+ \rightarrow \mathrm{Tr} AB \rho ~~;~~
   A^- B^- \rightarrow \mathrm{Tr} \rho AB~~;~~
   A^+ B^- \rightarrow \mathrm{Tr}A \rho B\,,\label{dict}\ee etc. Therefore from the cyclic property of the trace the expectation value of the (ALP) field in the total density matrix is
\be  \langle a(\vx,t) \rangle   =   \mathrm{Tr}a^+(\vx,t)\,\hat{\rho}(0)=   \mathrm{Tr} \hat{\rho}(0)\,a^-(\vx,t)=  \mathrm{Tr} \A(\vx,t)\,\hat{\rho}(0) = \overline{\langle \langle \A(\vx,t) \rangle \rangle} \,,\label{averageA} \ee
whereas
\be \mathrm{Tr} \R(\vx,t)\,\hat{\rho}(0) =0 \label{aveR}\,. \ee
 We now introduce
\be \mathcal{C}^>_k(t,t') = \mathrm{Tr} a^{-}_{\vk}(t) a^{+}_{-\vk}(t') \hat{\rho}(0)~~;~~ \mathcal{C}^<_k(t,t') = \mathrm{Tr} a^{-}_{\vk}(t') a^{+}_{-\vk}(t) \hat{\rho}(0)\,, \label{gis}\ee and the energy per mode of wavevector $\vk$
\be \mathcal{E}_k = \frac{1}{4}\,  \Bigg(\frac{\partial}{\partial t}\frac{\partial}{\partial t'} +\Omega^2_k  \Bigg)\Bigg[\mathcal{C}^>_k(t,t')+\mathcal{C}^<_k(t,t') \Bigg]_{t=t'} \,, \label{energy} \ee where we anticipate a renormalization of the frequency $\omega_k \rightarrow \Omega_k$,   which will be addressed in detail below. Using the definition (\ref{kelvars}) and the relations (\ref{dict}) it is straightforward to show that this symmetrized product yields
\bea\mathcal{E}_k & = &  \frac{1}{2 } \mathrm{Tr} \Bigg(\dot{\mathcal{A}}_{\vk}(t)\dot{\mathcal{A}}_{-\vk}(t)+ \Omega^2_k {\mathcal{A}}_{\vk}(t) {\mathcal{A}}_{-\vk}(t)  \Bigg)\,\hat{\rho}(0) \nonumber \\ & = & \frac{1}{2}\,\Bigg\{\overline{\langle \langle \dot{\mathcal{A}}_{\vk}(t)\dot{\mathcal{A}}_{-\vk}(t) \rangle \rangle}+ \Omega^2_k \,\overline{\langle \langle {\mathcal{A}}_{\vk}(t) {\mathcal{A}}_{-\vk}(t) \rangle \rangle}\Bigg\} \,.\label{eneA} \eea which  is the average energy per mode, a component of the energy momentum tensor. This analysis confirms that at least for the time evolution of the expectation values of the (ALP) field and its energy (momentum tensor) only the center of mass Keldysh variable $\A$ is needed.

\subsection{General properties of environmental correlation functions:}\label{subsec:corre}
The dynamics and dissipative processes depend on the correlation functions of the environment and crucially on their spectral density, these correlation functions determine the self-energy $\Sigma$ and the noise correlation function $\mathcal{N}$.

Because the bath is in thermal equilibrium, its  initial density matrix is $\rho_\chi(0)=e^{-\beta H_{\chi}}/Tr\, e^{-\beta H_{\chi}}$ which is space-time translationally  invariant, and the Heisenberg picture operators associated with the bath are given by $\mathcal{O}_\chi(\vx,t) = e^{iH_{\chi}t}\,\mathcal{O}_\chi(\vx,0)\,e^{-iH_{\chi}t}$ we can write
\bea  G^>(\vx-\vx';t-t')  & = &  \langle \mathcal{O}_\chi(\vx,t)\mathcal{O}_\chi(\vx',t') \rangle_\chi = \int \frac{d^4k}{(2\pi)^4}~ \rho^>(\vk,k_0) e^{-ik_0(t-t')}\,e^{i\vk\cdot(\vx-\vx')} \label{Ggfd} \\
G^<(\vx-\vx';t-t')  & = &  \langle \mathcal{O}_\chi(\vx',t')\mathcal{O}_\chi(\vx,t) \rangle_\chi = \int \frac{d^4k}{(2\pi)^4}~ \rho^<(\vk,k_0) e^{-ik_0(t-t')}\,e^{i\vk\cdot(\vx-\vx')} \,. \label{Glfd} \eea These representations are obtained by  writing $\op(\vx,t) = e^{iH_\chi t}\,e^{-i\vec{P}\cdot \vx} \,\op(\vec{0},0) \,e^{-iH_\chi t}\,e^{i\vec{P}\cdot \vx}$ and introducing a complete set of simultaneous eigenstates of $H_{\chi}$ and the total momentum operator $\vec{P}$,  $(H_\chi,\vec{P})\ket{n} = (E_n,\vec{P}_n)\ket{n}$, from which we obtain the following Lehmann representations,
\begin{eqnarray}
\rho^>(k_0,\vk) & = &  \frac{(2\pi)^4}{\mathrm{Tr}\rho_\chi(0)}~
\sum_{m,n}e^{-\beta E_n}
|\langle n| {\cal O}_\chi(\vec{0},0) |m \rangle|^2  \, \delta(k_0-(E_m-E_n))\,\delta(\vec{k}-(P_m-P_n)) \label{siggreat} \\
\rho^<(k_0,\vk) & = &  \frac{(2\pi)^4}{\mathrm{Tr}\rho_\chi(0)}~
\sum_{m,n} e^{-\beta E_n}
 |\langle m| {\cal O}_\chi(\vec{0},0) |n \rangle|^2  \, \delta(k_0-(E_n-E_m))\,\delta(\vec{k}-(P_n-P_m))\,.
 \label{sigless}
\end{eqnarray} Upon relabelling
$m \leftrightarrow n$ in the sum in the definition (\ref{sigless}) and recalling that $\mathcal{O}$ is an hermitian operator,
we find the Kubo-Martin-Schwinger relation\cite{kms,kapusta,lebellac,dasbuk,das2}

\begin{equation}
\rho^<(k_0,k)  = \rho^>(-k_0,k) = e^{-\beta k_0}
\rho^>(k_0,k)\,. \label{KMS}
\end{equation}

  The spectral density is defined as
\be \rho(k_0,k) = \rho^>(k_0,k)-\rho^<(k_0,k) = \rho^>(k_0,k)\big[ 1-e^{-\beta k_0}\big] \label{specOs}\ee
therefore
\be  \rho^>(k_0,k) = \rho(k_0,k)~\big[1+n(k_0)\big]~~;~~\rho^<(k_0,k) = \rho(k_0,k)~ n(k_0) \,, \label{relas}\ee  where
\be n(k_0) = \frac{1}{e^{\beta k_0}-1} \,. \label{bose}\ee

Furthermore, from the first equality in (\ref{KMS}) it follows that
\bea \rho(-k_0,k) & = &  - \rho(k_0,k) \,,  \label{oddros}\\ \rho(k_0,k) & > & 0 ~~\mathrm{for} ~~ k_0 > 0 \,. \label{positive}
\eea

In terms of the spectral densities we find
\be  \Big[G^>(\vx-\vx';t-t')-G^<(\vx-\vx';t-t')\Big]   = \int \frac{d^4k}{(2\pi)^4}\,\rho(k_0,k)  e^{-ik_0(t-t')}\,e^{i\vk\cdot(\vx-\vx')} \label{Godd} \ee which determines the self-energy $\Sigma(t-t')$ eqn. (\ref{kernelsigma}), and
\be    \Big[G^>(\vx-\vx';t-t')+G^<(\vx-\vx';t-t')\Big]   \equiv      \int \frac{d^4k}{(2\pi)^4}\,\widetilde{\mathcal{K}}(k_0,k)  e^{-ik_0(t-t')}\,e^{i\vk\cdot(\vx-\vx')} \label{Geven} \ee which determines the
noise correlation function $\mathcal{N}(t-t')$, eqn. (\ref{kernelkappa}), where
\be \widetilde{\mathcal{K}}(k_0,k) =  \rho(k_0,k)\,\coth\big[ \frac{\beta k_0}{2}\big]\,,\label{kernelK}\ee

Equation (\ref{kernelK}) is the general form of the fluctuation dissipation relation. Note that $\rho(k_0,k)$ is \emph{odd}   whereas $\widetilde{\mathcal{K}}(k_0,k)$ is \emph{even} in $k_0$. We emphasize that these are exact relations, the ``environmental'' fields $\chi$ may be coupled to other fields, for example, in the case of the (ALP) interaction with the electromagnetic fields as in eqn. (\ref{inter}) the gauge field also interacts with electrons, charged leptons and quarks, and similarly with the possible interaction with fermionic fields in eqn. (\ref{inter}), these interact with other gauge fields. The relations (\ref{KMS}-\ref{kernelK}) are general, non-perturbative statements relying on thermal equilibrium and space-time translational invariance and do not depend on these couplings.

The general expressions (\ref{Godd},\ref{Geven}) allow us to write the self-energy $\Sigma_k(t-t')$ (\ref{kernelsigma}) and the noise correlation function $\mathcal{N}_k(t-t')$ (\ref{kernelkappa}) as
\bea \Sigma_k(t-t')& = & -ig^2\int \frac{dk_0}{(2\pi)}\,\rho(k_0,k)  e^{-ik_0(t-t')}  \label{sigmadis} \\ \mathcal{N}_k(t-t') & = & \frac{g^2}{2}\,\int \frac{dk_0}{(2\pi)}\,\rho(k_0,k)\, \coth\big[ \frac{\beta k_0}{2}\big]\,  e^{-ik_0(t-t')} \,, \label{noisedis} \eea this is the general relation between the self-energy and the noise correlation function commonly determined by the spectral density $\rho(k_0,k)$, a direct consequence of the fluctuation-dissipation relation as a result of the bath being in thermal equilibrium.

\subsection{Misaligned initial conditions:}\label{subsec:misa}

The initial density matrix for the (ALP) field   is determined by  initial conditions. We consider an initial density matrix describing a pure state compatible with a ``misalignment'' mechanism whereby the expectation value of the (ALP) field is non-vanishing initially and also allow a non-vanishing expectation value of its canonical momentum. This is achieved by considering a \emph{coherent state} of the form
\be \ket{\Delta} = \Pi_{\vk} e^{\Delta_{\vk}\,b^\dagger_{\vk}-\Delta^*_{\vk} b_{\vk}}\,\ket{0}\,, \label{cs} \ee where $\ket{0}$ is the free field (ALP) vacuum state, $b^\dagger_{\vk}, b_{\vk}$ are (ALP) free field creation and annihilation operators, and $\Delta_{\vk}$ are complex c-number coefficients that determine the initial values for $\A_k,\pi_k$. In the Schroedinger representation the state (\ref{cs}) is represented by the coherent state wavefunctional
\be \Psi[a] =   e^{i\int d^3 x \overline{\pi}_i(\vx)a(x)}\,\Psi_0[a-\overline{\mathcal{A}}_i]\,, \label{psini}\ee where $\Psi_0$ is the ground state wavefunctional of a free (ALP)  field theory. Such wavefunctional is   Gaussian and yields an average momentum $\overline{\pi}_i$ and expectation value of the field given by $ {\overline{\A}}_i$ whose Fourier expansion is determined by the complex coefficients  $\Delta_{\vk}$ in eqn. (\ref{cs}). The pure state density matrix describing this coherent state as representative of the ``misaligned'' initial condition is
\be \rho_a[a,a';0] =  \Psi^*[a']\Psi[a]\,,  \label{rhomisi}\ee and its Wigner transform is given by
\be W[\A_i,\pi_i] = N\,\Pi_{\vk}\, e^{-\frac{\Omega_k}{2}(\A_{i,\vk}-\overline{\A}_{i,\vk})(\A_{i,-\vk}-\overline{\A}_{i,-\vk})} \,e^{-\frac{1}{2\Omega_k}(\pi_{i,\vk}-\overline{\pi}_{i,\vk})(\pi_{i,-\vk}-\overline{\pi}_{i,-\vk})}\,, \label{Wofapi}\ee with $N$ a normalization factor and $\Omega_k$ the covariance which will be related to the renormalized effective frequency (see below).
 Translational invariance imposes that the expectation value of the (pseudo) scalar (ALP) field be independent of the momentum, therefore we write in a finite but large quantization volume $V$
   \be \overline{\A}_{i,\vk} = \overline{\A}_i \sqrt{V}\,\delta_{\vk,0}~~;~~ \overline{\pi}_{i,\vk} = \overline{\pi}_i \sqrt{V}\,\delta_{\vk,0}\,, \label{traninvA}\ee where $\overline{\A}_i;\overline{\pi}_i$ are the space-time constant expectation values of the field and canonical momentum in the translational invariant initial state. With this Wigner probability distribution function we find the averages over the initial conditions

  \bea && \overline{(\A_{i,\vk})}= \overline{\A}_i \sqrt{V}\,\delta_{\vk,0}~;~ \overline{(\A_{i,\vk}-\overline{\A}_{i,\vk})(\A_{i,-\vk}-\overline{\A}_{i,-\vk})}= \frac{1}{\sqrt{2\Omega_k}} \nonumber \\
   &&    \overline{(\pi_{i,\vk})}= \overline{\pi}_i \sqrt{V}\,\delta_{\vk,0}  ~;~
    \overline{(\pi_{i,\vk}-\overline{\pi}_{i,\vk})(\pi_{i,-\vk}-\overline{\pi}_{i,-\vk})}= \sqrt{\frac{\Omega_k}{2}}\,,\label{avewigners}  \eea with higher order correlations obtained via Wick's theorem.

 This is a simple realization of the ``misalignment'' mechanism whereby the initial state is a coherent state that features a non-vanishing expectation value of the field and its canonical momentum, these define the initial value problem.

\subsection{The solution of the Langevin equation}\label{subsec:solangevin}
The solution of the Langevin (stochastic) equation (\ref{langevin}) is obtained by Laplace transform,
define the Laplace transforms
\bea \widetilde{\A}_{\vk}(s) & = &  \int^\infty_0 e^{-st}\,\A_{\vk}(t)\,dt \,,\label{laplaA}    \\ \widetilde{\xi}_{\vk}(s) & = &  \int^\infty_0 e^{-st}\,\xi_{\vk}(t)\,dt \,,\label{laplachi} \\ \widetilde{\Sigma}_{\vk}(s) & = &  \int^\infty_0 e^{-st}\,\Sigma_{\vk}(t)\,dt  = -\frac{g^2}{2\pi} \,\int^{\infty}_{-\infty} \frac{\rho(k_0,k)}{k_0-is} dk_0 \,,\label{laplasigma}
\eea where in (\ref{laplasigma}) we used the dispersive representation (\ref{sigmadis}).

With the initial conditions (\ref{inicons}) the solution of the Laplace transform of the Langevin equation is
\be \widetilde{\A}_{\vk}(s) = \frac{\pi_{i,\vk}+s\,\A_{i,\vk}+\widetilde{\xi}_{\vk}(s)}{s^2+\omega^2_k+\widetilde{\Sigma}_{\vk}(s)}\,.\label{laplasolution}\ee
The solution in real time is obtained by inverse Laplace transform, it is given by
\be \A_{\vk}(t) = \A_{\vk;h}(t) +  \A_{\vk;\xi}(t)\,, \label{Asplits} \ee  where $\A_{\vk;h}; \A_{\vk;\xi}(t)$ are the homogeneous   and inhomogeneous solutions respectively, namely

\bea \A_{\vk;h}(t)  & =  & \A_{i,\vk}\,\dot{\mathcal{G}}_k(t) + \pi_{i,\vk}\,\mathcal{G}_k(t) \nonumber  \\   \A_{\vk;\xi}(t) & =  & \int^t_0 \mathcal{G}_k(t-t')\xi_{\vk}(t')\,dt' \,,\label{realtisol} \eea  and the Green's function is given by
\be \mathcal{G}_k(t) = \frac{1}{2\pi i} \int_{\mathcal{C}} \frac{e^{st}}{s^2+\omega^2_k+\widetilde{\Sigma}_{\vk}(s)}\, ds \,, \label{goftsol}\ee   $\mathcal{C}$ denotes the Bromwich contour parallel to the imaginary axis   and to the right of all the singularities   of $(s^2+\omega^2_k+\widetilde{\Sigma}_{\vk}(s))^{-1}$ in the complex s-plane and closing along a large semicircle at infinity with $Re(s)<0$. These singularities correspond to poles and multiparticle branch cuts with $Re(s)<0$, thus the contour runs parallel to the imaginary axis $s= i(\nu -i \epsilon)$, with $-\infty \leq \nu \leq \infty$ and $\epsilon \rightarrow 0^+$. Therefore,
\be \mathcal{G}_k(t) = - \int^{\infty}_{-\infty} \widetilde{\mathcal{G}}_k(\nu)\, {e^{i\nu\,t}} \,\frac{d\nu}{2\pi}\,, \label{Goftfin}\ee  where
\be \widetilde{\mathcal{G}}_k(\nu) = \frac{1}{(\nu-i\epsilon)^2 -\omega^2_k - \Sigma(\nu,k) }\,. \label{Gfnu}  \ee

The self energy in frequency space is given by the dispersive form
\be \Sigma (\nu,k)   =   \frac{g^2}{2\pi}\,\int^{\infty}_{-\infty}  \,  \frac{\rho(k_0,k)}{\nu-k_0-i\epsilon} \,dk_0 \equiv \Sigma_R(\nu,k) + i \Sigma_I(\nu,k) \,,
 \label{signu}\ee with the real and imaginary parts given by
\bea \Sigma_R(\nu,k) & = &  \frac{g^2}{2\pi}\,\mathcal{P}\int^{\infty}_{-\infty}   \Bigg[ \frac{\rho(k_0,k)}{\nu-k_0}\Bigg]\,dk_0 \,,\label{resig}\\
  \Sigma_I(\nu,k) & = & \frac{g^2}{2}\,\rho(\nu,k) \,, \label{imsig} \eea yielding the Kramers-Kronig relation
  \be \Sigma_R(\nu,k) = \frac{1}{\pi}\,\mathcal{P} \int^{\infty}_{-\infty}   \Bigg[ \frac{\Sigma_I(k_0,k)}{\nu-k_0}\Bigg]\,dk_0 \,. \label{KKrela}\ee
   To obtain the above representations we have used the relation $\rho(-k_0,k) = -\rho(k_0,k)$ (see eqn. (\ref{positive})), as a consequence of which it follows that $\Sigma_R(\nu,k) = \Sigma_R(-\nu,k)~;~\Sigma_I(\nu,k) = - \Sigma_I(-\nu,k)$. $\widetilde{\mathcal{G}}_k(\nu)$ given by eqn. (\ref{Gfnu})   features   complex poles corresponding to the solution of the equation
  \be \omega^2_P(k) = \omega^2_k  + \Sigma(\omega_P(k),k)\,, \label{poleG}\ee to leading order in $g^2$ we find
  \be \omega_P(k) = \pm \Omega_k + i \frac{\Gamma_k}{2}\,, \label{polval}\ee where
  \be \Omega_k = \omega_k + \frac{\Sigma_R(\omega_k,k)}{2\omega_k} ~~;~~ \Gamma_k = \frac{\Sigma_I(\omega_k,k)}{\omega_k}= \frac{g^2}{2\omega_k}\,\rho(\omega_k,k) \,. \label{reimpole}\ee Writing in the denominator of the integrand in (\ref{Goftfin}) $\Sigma(\nu,k) = \Sigma(\omega_p(k),k) + (\Sigma(\nu,k)-\Sigma(\omega_p(k),k))$ we find that near each pole, $\widetilde{\mathcal{G}}_k(\nu)$ can be written in
  a Breit-Wigner form as
  \be \widetilde{\mathcal{G}}_k(\nu) =  \frac{Z}{2\omega_P(k)(\nu \mp \Omega_k -i\frac{\Gamma_k}{2})} \,\,, \label{BW} \ee with the wave function renormalization constant
   \be Z^{-1} = 1- \frac{\Sigma'(\omega_P,k)}{2\omega_P} = 1 + \mathcal{O}(g^2)~,~~~ \Sigma'(\omega_P,k) \equiv \Big[\frac{d}{d\nu}\Sigma'(\nu,k)\Big]_{\nu=\omega_P}\,. \label{zwf} \ee To leading order in $g^2$ we find
   \be \mathcal{G}_k(t) = e^{-\frac{\Gamma_k}{2}t}\,\frac{\sin(\Omega_k t)}{\Omega_k } + \mathcal{O}(g^2)\,, \label{goftbw} \ee where we have assumed a narrow width $\Gamma_k/\Omega_k \propto g^2 \ll 1$ and neglected terms of this order. Using  this result in eqn. (\ref{realtisol}) we find

   \bea \overline{\langle \langle \mathcal{A}_{\vk}(t) \rangle\rangle}  & = &   e^{-\frac{\Gamma_k}{2}t}\,\Bigg\{  \overline{\mathcal{A}}_{i,\vk}    \Big[\cos(\Omega_k t) - \frac{\Gamma_k}{2\Omega_k}\,\sin(\Omega_k t) \Big] + \overline{\pi}_{i,\vk}\,\frac{\sin(\Omega_k t)}{\Omega_k} \Bigg\}+\mathcal{O}(g^2) \,,  \label{aveA} \\
   \overline{\langle \langle \dot{\mathcal{A}}_{\vk}(t) \rangle\rangle}  & = &   e^{-\frac{\Gamma_k}{2}t}\,\Omega_k\Bigg\{  \overline{\mathcal{A}}_{i,\vk}    \Big[-\sin(\Omega_k t) - \frac{\Gamma_k}{2\Omega_k}\,\cos(\Omega_k t) \Big] + \overline{\pi}_{i,\vk}\, \frac{\cos(\Omega_k t)}{\Omega_k} \Bigg\}  \nonumber \\ & & -\frac{\Gamma_k}{2}\overline{\langle \langle \mathcal{A}_{\vk}(t) \rangle\rangle} + \mathcal{O}(g^2)  \label{avedotA}
   \eea where we used (\ref{noiscors}), and $\overline{A}_i,\overline{\pi}_i$ are the average of the initial conditions with the Wigner distribution function (\ref{avewigners}). We have explicitly displayed the terms $\propto \Gamma_k/\Omega_k$ to exhibit that they arise from the derivative of the exponential damping term, however, these terms are of $\mathcal{O}(g^2)$ and must be neglected for consistency as we are also neglecting terms of the same order from wave function renormalization.

    Similarly, we find
   \be \overline{\langle \langle \mathcal{A}_{\vk}(t) \mathcal{A}_{-\vk}(t)\rangle\rangle} =   \overline{\mathcal{A}_{\vk;h}(t) \mathcal{A}_{-\vk;h}(t)}   + \frac{g^2}{4\pi} \int^{\infty}_{-\infty} \rho(k_0,k) \coth\big[\frac{\beta k_0}{2}\big] \,\Big| \int^t_0 \mathcal{G}_k(\tau)\,e^{ik_0\tau}\,d\tau \Big|^2 \,dk_0 \,.\label{Asq} \ee
   Using the leading order result (\ref{goftbw}) for $\mathcal{G}_k(\tau)$ the integral in (\ref{Asq}) is straightforward. Inserting the result into (\ref{Asq}) yields four terms, the resulting integrals are performed by contour integration in the complex $k_0$-plane: in the narrow width approximation the two direct terms feature residues $\propto 1/\Gamma_k$, whereas the interference terms feature residues $\propto 1/(2\Omega_k+i\Gamma_k)$, these latter terms and  the poles at $k_0 = 2\pi\,i m/\beta, m = 0, \pm 1, \pm 2 \cdots$, namely the Matsubara frequencies\footnote{The residue for $m=0$ vanishes because the spectral density vanishes at $k_0=0$. }, yield contributions of $\mathcal{O}(g^2)$ and will be neglected, whereas the terms with residues $\propto 1/\Gamma_k \propto 1/g^2$ yield the leading contributions. Using the definition of $\Gamma_k$ (\ref{reimpole}) and keeping solely the leading order terms in (\ref{aveA}) we obtain
   \be \overline{\langle \langle \mathcal{A} _{\vk}(t)\mathcal{A} _{-\vk}(t)  \rangle\rangle} = e^{-\Gamma_k t}\,\,\overline{\mathcal{H}_{\vk}(t)\,\mathcal{H}_{-\vk}(t)}+ \frac{1}{2\Omega_k}\,\Big[ 1 + 2 n(\Omega_k)\Big]\, \Big(1-e^{-\Gamma_k t}\Big)+ \mathcal{O}(g^2)\,, \label{asquareave} \ee where
   \be \mathcal{H}_{\vk}(t) = \mathcal{A}_{i,\vk}\cos(\Omega_k t) +  {\pi}_{i,\vk}\,\frac{\sin(\Omega_k t)}{\Omega_k}\,, \label{ache}   \ee
    and  $n(\Omega_k)$ is the Bose-Einstein distribution function.

   This is a noteworthy result:  for $\Gamma_k t \gg 1$  the surviving term is precisely the free field expectation value $ (\langle b^\dagger_{\vk} b_{\vk}+b_{\vk} b^\dagger_{\vk} \rangle)/2\Omega_k$ of (ALP) operators,  where the average is in a thermal equilibrium statistical ensemble, namely at long time $t \gg 1/\Gamma_k$ the (ALP) particles \emph{thermalize} with the bath.

    A similar calculation, implementing the same approximations yields for the average energy per mode (\ref{eneA})
   \be \mathcal{E}_k = \frac{e^{-\Gamma_k t}}{2}\, \Big[\, \overline{\dot{\mathcal{H}}_{\vk}(t)\,\dot{\mathcal{H}}_{-\vk}(t)}+ \Omega^2_k \, \overline{\mathcal{H}_{\vk}(t)\,\mathcal{H}_{-\vk}(t)}\,  \Big]+ \frac{\Omega_k}{2} \,\Big[1+ 2 n(\Omega_k)\Big]\,\Big(1-e^{-\Gamma_k t}\Big)+ \mathcal{O}(g^2)\,.\label{eneasy} \ee

   This result  confirms thermalization at long time, the second term, which survives for $t\gg 1/\Gamma_k$ is identified as the expectation value of the free field Hamiltonian in a thermal density matrix, namely the  internal energy. These results are a manifestation of thermalization in the same manner as a Brownian oscillator as mentioned in the introduction: whereas the average of the coordinate relaxes to the minimum of the potential, the mean square root fluctuations reveal thermalization with the bath. This is ultimately a consequence of the fluctuation-dissipation relation manifest in the relation (\ref{kernelK}) between the noise   and the self-energy (friction) kernels,  a corollary of the Kubo-Martin-Schwinger condition (\ref{KMS}) as a consequence of the equilibrium bath correlations.

     Therefore, for the Wigner distribution function (\ref{Wofapi}) describing a misaligned initial condition with the averages given by eqns. (\ref{avewigners}) and neglecting a zero point contribution,   we find the  energy density
      \be \frac{E}{V} = \frac{1}{V} \sum_{\vk} \mathcal{E}_k =   \frac{e^{-\Gamma_0 t}}{2} \Big[\overline{\pi}^2_i + m^2_a\,\overline{\A}^2_i\Big]+  \int \frac{d^3 k}{(2\pi)^3}\, \Omega_k \,  n(\Omega_k) \,\Big(1-e^{-\Gamma_k t}\Big)+ \mathcal{O}(g^2)\,.  \label{enerdens}\ee
      The first term is identified with a \emph{cold} dark matter contribution and originates in the damped coherent oscillations arising from a ``misaligned'' initial condition as in the usual case of an (ALP), whereas the second term yields a \emph{hot} dark matter contribution from the approach to thermalization, each weighted by the damping exponentials. Whereas the first term depends on the initial conditions of the (ALP), the second term is completely determined by the noise, namely the thermal bath. This is one of the important results of this study.

       Therefore, if the (ALP) relaxes on cosmological time scales at a given time $t$ its contribution to dark matter is a \emph{mixture} of cold and hot components, with a fraction determined by the relaxation rate $\Gamma_k$ and the time scale $t$. This result suggests a scenario where the ``warmth'' of the dark matter evolves in time from colder to hotter.

       The   result (\ref{enerdens}) is general, it is valid to   order   $g^2$ for any (ALP) interaction of the form $g a(x) \mathcal{O}(x)$  and to all orders in the interactions of the bath fields with other fields besides the (ALP). This is an important corollary of the results in this section: the processes that yield the friction in the equation of motion of the misaligned expectation value are the \emph{same} processes that lead to thermalization. Therefore, whereas several studies focused on the friction term in the equation of motion of the coherent condensate\cite{mottola,friction,friction1} and other studies focused on thermalization\cite{buch,masso} our results show that both processes are related by the fluctuation dissipation relation, occur on similar time scales and both contribute to the   evolution of the energy density of the (ALP) field. Therefore, the  time evolution of the energy density given by eqn. (\ref{enerdens}) is one of the important results of our study, it  applies to all dissipative processes resulting from interactions of the ALP with other degrees of freedom and is a direct consequence of the fluctuation dissipation relation.

   \section{ALP interacting with photons:}\label{sec:EBint}
   The results obtained in the previous section are general, although the focus is on (ALP) fields, the results also apply to any field with an interaction of the form  (\ref{lag}) and initial conditions that allow for the evolution of a coherent condensate\cite{turner}. These results have a clear physical significance in terms of the non-equilibrium manifestation of Brownian fluctuations: a bath in equilibrium induces both a self-energy (friction) and a noise term in the effective equations of motion, the spectral
   properties of both are related by the generalized fluctuation dissipation relation, a hallmark of a bath in thermal equilibrium. Although the results are general, the details, namely relaxation times, frequency renormalization etc. depend on the spectral properties of the bath correlations. In this section we focus on (ALP) interaction with photons via the coupling
   \be \mathcal{L}_I = -g a(x) \vec{E}(x)\cdot\vec{B}(x) \,,\label{aeb} \ee as in eqn. (\ref{inter}). The main assumption invoked in our study is that we consider free \emph{massless} photons neglecting interactions with charged leptons and quarks.
   The one loop contribution to the (ALP) self-energy is displayed in fig. (\ref{fig:se}).

       \begin{figure}[h!]
\includegraphics[height=3.5in,width=3in,keepaspectratio=true]{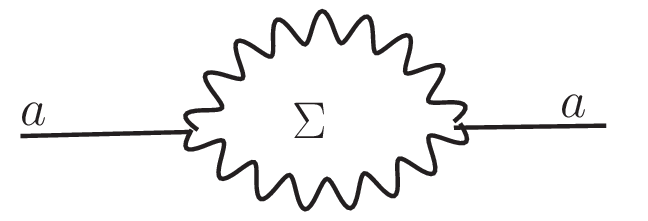}
\caption{One loop (ALP) self-energy from the coupling $\mathcal{L}_I = -g a(x) \vec{E}(x)\cdot\vec{B}(x)$.}
\label{fig:se}
\end{figure}

   The regime of validity of this assumption is discussed in detail in section (\ref{sec:discussion}) below. An important aspect of the coupling (\ref{aeb}) is that this interaction is non-renormalizable because the coupling $g$ has dimensions of $1/\mathrm{energy}$. As a result the loop corrections associated with the self-energy feature ultraviolet divergences which cannot be absorbed into the parameters of the Lagrangian and the theory must be interpreted as a low energy effective field theory.

   In appendix (\ref{app:ebcoup}) we obtain the spectral density from the thermal correlation functions of the composite operator $ \vec{E}(x)\cdot\vec{B}(x)$, it is given by (see eqn. (\ref{rhofi})),
\bea
  &&  \rho(q_0,\vec{q})
     =   \frac{(Q^2)^2}{32\pi}\,\Bigg\{\Bigg(1 + \frac{2}{\beta q}\,\ln\Bigg[\frac{1-e^{-\beta \omega^I_+}}{1-e^{-\beta \omega^I_-}} \Bigg]\Bigg)\,\Theta(Q^2)  + \frac{2}{\beta q}\, \ln\Bigg[\frac{1-e^{-\beta \omega^{II}_+}}{1-e^{-\beta \omega^{II}_-}} \Bigg]\,\Theta(-Q^2) \Bigg\}\, \mathrm{sign}(q_0)\,, \nonumber  \\  &&  Q^2 = q^2_0-q^2 ~~;~~  \omega_\pm^{(I)}   =   \frac{|q_0| \pm q}{2}~~;~~
    {\omega}_\pm^{(II)} = \frac{q \pm |q_0|}{2}\,. \label{rhofiI} \eea The terms with $\Theta(Q^2)$ arise from the processes $a \leftrightarrow 2 \gamma$, namely emission and absorption of photons with the reverse or recombination process $2\gamma \rightarrow a$ a consequence of the heat bath, these processes  feature support on the (ALP) mass shell for massive (ALP) particles. The contribution proportional to $\Theta(-Q^2)$ only features support below the light cone and describes off-shell processes $\gamma a \leftrightarrow \gamma$.
    This interpretation stems from the delta functions in the expressions for the spectral density eqn. (see the second line in eqn. (\ref{rhoeb})).

    From the definition of the relaxation rate (\ref{reimpole}) and with the result (\ref{rhofiI}) we find
    \be \Gamma_T = \Gamma \, \Bigg(1 + \frac{2}{\beta q}\,\ln\Bigg[\frac{1-e^{-\beta \omega^I_+}}{1-e^{-\beta \omega^I_-}} \Bigg]\Bigg)_{k_0=\Omega_k}~~;~~ \Gamma = \frac{g^2\,m^4_a}{64\pi\,\Omega_k}\,, \label{gammaeb} \ee the first contribution is the zero temperature (ALP) decay rate, and the second is the finite temperature contribution  which is a consequence of stimulated emission and absorption in the heat bath. The ratio $\Gamma_T/\Gamma$ is displayed in fig. (\ref{fig:gamaT}) as a function of the dimensionless ratios $T/m_a,k/m_a$.

     \begin{figure}[h!]
\includegraphics[height=3.5in,width=3.5in,keepaspectratio=true]{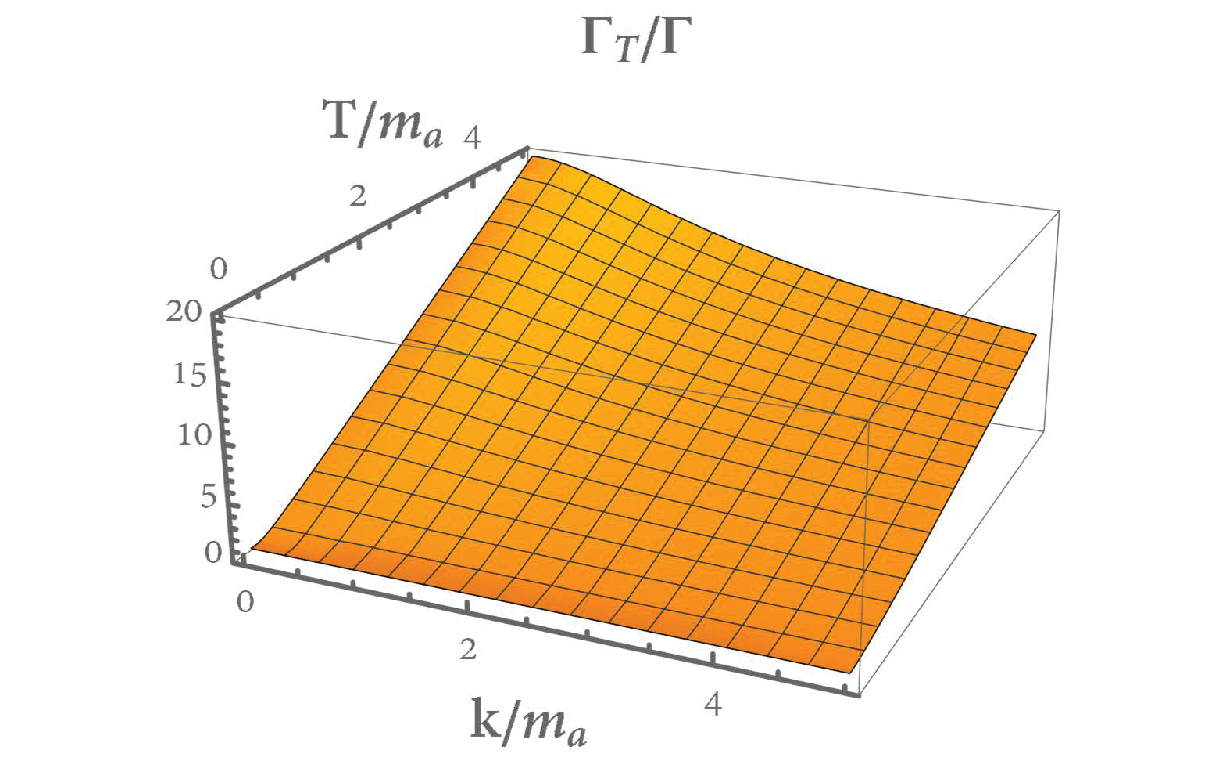}
\caption{Ratio $\Gamma_T/\Gamma$ vs $T/m_a,k/m_a$.}
\label{fig:gamaT}
\end{figure}

     The finite temperature contribution  yields a large enhancement over the zero temperature case  for  $T\gg m_a, k$. For example in the long-wavelength limit $k\ll m_a$ we find
    \be \Gamma_T = \frac{g^2\,m^3_a}{64\pi} \, \Bigg(1 +  {2}\,n\Big(\frac{m_a}{2}\Big)  \Bigg)\,, \label{kzerog}\ee which in the high temperature limit $T \gg m_a$ yields
    \be  \Gamma_T = \frac{g^2\,m^3_a}{16\pi} \Bigg( \frac{T}{m_a}\Bigg)\,. \label{hiTg}\ee For example, if $T$ corresponds to the temperature of the cosmic microwave background today $T \simeq 10^{-4}\,\mathrm{eV}$ the finite temperature correction yields a large enhancement for $m_a \ll \mu eV$, and an enormous one if the (ALP) is an ultra-light candidate with $m_a \lesssim 10^{-22}\,\mathrm{eV}$ with potentially relevant cosmological consequences discussed below in section (\ref{sec:discussion}).

    \vspace{1mm}

    \textbf{Real part of the self-energy: $i:)\, T=0$:}
The real part of the self energy is given by eqn. (\ref{resig}),  with the zero temperature contribution to the spectral density (\ref{rhofiI})
we find
\be \Sigma^{(0)}_R(\nu,k) = \frac{g^2}{64\pi^2} \mathcal{P} \Bigg\{ \int_0^\infty \frac{(k_0^2 - k^2)^2}{\nu - k_0} \Theta(k^2_0-k^2) dk_0 - \int_{-\infty}^0 \frac{(k_0^2 - k^2)^2}{\nu - k_0} \Theta(k^2_0-k^2) dk_0 \Bigg\}\,, \label{sigT0} \ee relabelling $k_0 \rightarrow -k_0$ in the second integral, the total integral is ultraviolet divergent, introducing an upper frequency cutoff $\Lambda$ delimiting the range of validity of the effective (ALP) field theory,   and changing integration variables to $\alpha = k^2_0 - k^2 - (\nu^2-k^2)$, we find
\be \Sigma^{(0)}_R(\nu,k) = - \frac{g^2}{64\pi^2}   \int_{-K^2}^{\Lambda^2} \mathcal{P}\Bigg\{\alpha + 2K^2 + \frac{(K^2)^2}{\alpha}\Bigg\} d\alpha \, ~~;~~ K^2 = \nu^2-k^2 \,,  \label{sigT02}  \ee with the result
\be  \Sigma^{(0)}_R(\nu,k) = - \frac{g^2}{64\pi^2}  \Big[  \frac{1}{2}  \Lambda^4  + 2K^2 \Lambda^2  + \frac{3}{2}(K^2)^2 + (K^2)^2\, \ln\Big[\frac{\Lambda^2}{|K^2|}\Big]\Big]\,. \label{sigcero}
\ee This result clearly exhibits the non-renormalizability of the effective field theory of (ALP): the $\Lambda^4$ term is absorbed into a mass renormalization, the $(\nu^2-k^2)\,\Lambda^2$ term yields an ultraviolet divergent wave function renormalization as per eqn. (\ref{zwf}), however the logarithmic divergence $(\nu^2-k^2)^2\,\ln[\Lambda]$ cannot be absorbed into the renormalization of parameters and field redefinitions of the original Lagrangian, which then must be appended with a new higher derivative term $C (\partial_\mu \partial^\mu \,a)^2$ where $C$ is a new coefficient that will be renormalized by the term with the logarithmic divergence.  Therefore, the effective action necessitates the addition of a higher derivative term to absorb the ultraviolet divergences. While such   extension of the effective field theory  is both  necessary and interesting on its own,   here we focus on the minimal (ALP) effective field theory to establish contact with the more familiar (ALP) Lagrangians, thereby we set the new renormalized coupling $C = 0$. In section (\ref{sec:discussion}) we comment on possible effects associated with the higher derivative terms.

\textbf{ii:)\, $T\neq 0$: }
 The finite temperature contribution to the self-energy is ultraviolet finite and is studied in detail in appendix (\ref{app:finiTsig}), the results of this appendix allow to obtain its high and low temperature behavior. For $T \gg \Omega_k$ we find to leading orders in the high temperature expansion
\be {\Sigma^T_R(\Omega_k,k)}  =  {g^2 T^4}   \Bigg[ -\frac{\pi^2}{15} -  \frac{m^2_a}{24\,T^2}+ \frac{m^4_a }{32\,T^4} \, \bigg( 1 - \gamma + \ln\bigg[\frac{4\pi T}{m_a}\bigg] - \frac{\Omega_{\vec{k}}}{k} \ln\bigg[\frac{\Omega_{\vec{k}} + k}{m_a}\bigg] \bigg)
 +\cdots \Bigg]\,.\label{hiT} \ee and for $T \ll  m_a$ we find
\be \Sigma^T_R(\Omega_k,k) =   g^2 T^4  \,  \Bigg[\frac{4\pi^2}{45} \frac{k^2 }{m^2_a\,}+ \frac{32\pi^4\,m_a}{63}\Big(1 + 4 \frac{k^2}{m^2_a}+ \frac{16}{5} \frac{k^4}{m^4_a}  \Big)\,\frac{T^2}{m^2_a}+ \cdots  \Bigg] \,.\label{loT} \ee

 Defining the effective finite temperature mass as the $k\rightarrow 0$ limit of the dispersion relation  (\ref{poleG}), the high temperature limit $T \gg m_a$ (\ref{hiT}) yields an effective, temperature dependent mass
 \be m^2_a(T) = m^2_{aR} \Big[1- \frac{\pi^2\,g^2\,T^4}{15\,m^2_{aR}} \Big] \,, \label{m2ofT}\ee where $m_{aR}$ is the renormalized mass absorbing the zero temperature renormalization. Equation (\ref{m2ofT}) can be written in a more illuminating form as
 \be m^2_a(T) = m^2_a(0)\Big[1- \Big( \frac{T}{T_c}\Big)^4 \Big] ~~;~~ T_c = 1.11\sqrt{  \frac{m_a(0)}{g}} \,. \label{Tcdef}\ee

  This result \emph{suggests} the possibility of an \emph{inverted phase transition} at a temperature $T =T_c$: for $T>T_c$ the effective squared mass is \emph{negative} signalling an instability, whereas it is positive for $T<T_c$. This situation is the opposite of the usual phase transition where $m^2(T) > 0$  for $T>T_c$ indicating an ordered phase and symmetry restoration, and $m^2(T) <0$ for $T<T_c$ indicating symmetry breaking. This intriguing result is a consequence of the high temperature behavior of the real part of the self-energy, which to the best of our knowledge has not been studied before.

For $T <  T_c$, we define the finite temperature correction to the dispersion relation (\ref{reimpole}) as
\be \omega_T-\omega_a \equiv \frac{\Sigma^T(\Omega_k,k)}{2\Omega_k}\,, \label{def}\ee   in fig. (\ref{fig:disp}) we display the finite temperature correction to the dispersion relation  $(\omega_T-\omega_a)/g^2$ in units of $m^3_a$ vs. $T/m_a,k/m_a$.

     \begin{figure}[h!]
\includegraphics[height=3.5in,width=3.5in,keepaspectratio=true]{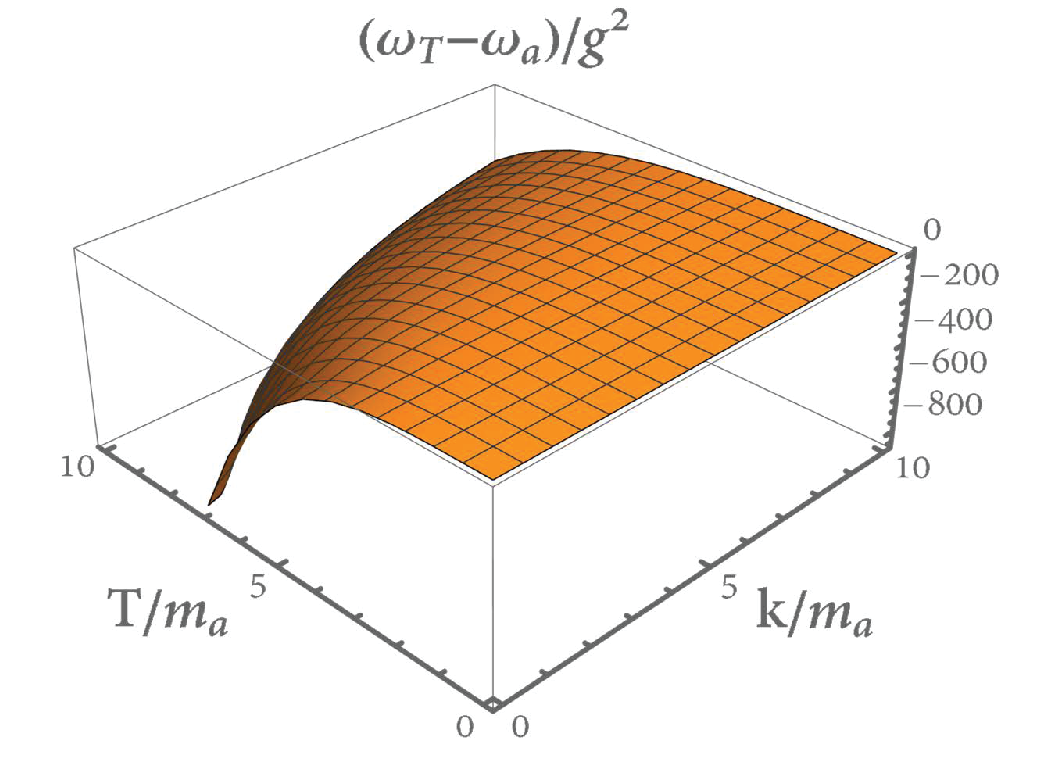}
\caption{Finite temperature correction to the dispersion relation  $(\omega_T-\omega_a)/g^2$ in units of $m^3_a$ vs $T/m_a,k/m_a$.}
\label{fig:disp}
\end{figure}

 For $k=0$ the figure clearly shows the fast drop in the effective mass  as $T$ increases in agreement with the analysis yielding eqn. (\ref{m2ofT}).

 The possible high temperature instability as a consequence of $m^2_a(T) < 0$ for $T> T_c$ indicates that the results obtained in the previous section for the energy density (\ref{enerdens}) are valid \emph{only} for $T < T_c$ since the solution of the Green's function (\ref{goftbw}) implied real frequencies $\Omega_k$  and a perturbative correction to the position of the poles. The instability for $T>T_c$ yields an imaginary frequency $\Omega_k$ in the solution which translates into a growing exponential.

   \section{Discussion and caveats:}\label{sec:discussion}

   \begin{itemize}

   \item \textbf{On the Gaussianity of noise correlations:} The noise variable $\xi$ is described by a Gaussian probability distribution function (PDF) given by eqn. (\ref{noispdf}). The Gaussianity is a consequence of the non-equilibrium effective action of the  (ALP) field being quadratic\cite{beilok}. However, this Gaussian (PDF) does not entail that either the (ALP) or the environmental fields are free. As per the discussion in section (\ref{subsec:corre}) the Lehmann representation of the environmental correlation functions is to all orders in the couplings of the environmental fields to other fields within or beyond the standard model \emph{other than} the (ALP) field. The fluctuation dissipation relation (\ref{kernelK}) is a consequence of the Lehmann representation, the self energy (\ref{sigmadis},\ref{laplasigma})), which enters in the full propagator (\ref{Gfnu})  is the sum  (to all orders) of one particle irreducible diagrams such as those displayed in fig. (\ref{fig:moreloops}). Therefore the spectral density $\rho(k_0,k)$ that determines the noise correlation functions   are also to all orders in such couplings. Hence, even when the noise (PDF) is Gaussian, this does not entail that either the (ALP) or the environmental fields are free.

  \item \textbf{(ALP) coupled to photons: region of validity.} In the case of (ALP) interaction with photons, we have assumed that photons constitute a thermal bath of blackbody radiation, having in mind the cosmic microwave background (CMB). At high temperature for relativistic electrons  namely $T \gg m_e$ with $m_e$ the electron mass, photons acquire a gauge invariant plasma mass  $\simeq eT/3$ via hard thermal loop corrections to the photon self-energy\cite{htl,kapusta,lebellac}. For a light or ultralight (ALP) this plasma mass would shut off the lowest order emission/absorption channel $a \leftrightarrow 2 \gamma$. When electrons become non-relativistic, but there is a free electron density $n$, the plasma frequency becomes $\Big(\frac{4\pi n e^2}{m_e}\Big)^{1/2}$  which would also shut off this channel for light or ultralight (ALP). However, after recombination, the free electron density vanishes precipitously as electrons combine with protons into neutral hydrogen. Photons are effectively massless as evidenced by the nearly perfect blackbody spectrum of the  (CMB). Since we have assumed massless photons in the calculation of the spectral density, our assumptions are valid after recombination for $T \lesssim 0.3\,\mathrm{eV}$. For a light (ALP) with $m_a \lesssim \,\mu eV$ even the temperature of the (CMB) today $T_{cmb} \simeq 10^{-4}\,eV$ is such that $T/m_a \gg 1$ and there is a large finite temperature enhancements to the relaxation rate, which becomes quite substantial for an ultralight (ALP) with $m_a \lesssim 10^{-20} eV$.

      \item \textbf{Thermalization:} Thermalization of (ALP) in the early Universe has been  studied previously\cite{turnerther,masso,buch}. However, our method and results go much further. The non-equilibrium effective action yields the effective equation of motion for (ALP) fields which is a Langevin equation with ``friction'' and noise contributions that satisfy the fluctuation dissipation relation. The solution of this Langevin equation allows us to study the evolution of (ALP) condensates from misaligned initial conditions  along with thermalization which is shown to be a consequence of the noise term and the fluctuation-dissipation relation. The effective action also allows us to study  renormalization aspects and the finite temperature corrections to the (ALP) mass arising from the real part of the self-energy (the thermalization rate is related to the imaginary part of the self-energy on the mass shell). The effective action has been obtained up to second order in (ALP) coupling, but \emph{to all orders} in the couplings of the ``environmental'' fields to any other fields to which they couple other than the (ALP).

          For example the study of thermalization in refs.\cite{turnerther,masso, buch} in which the (ALP) is coupled to quarks or other Standard Model degrees of freedom correspond to obtaining the \emph{two loop} contributions to $G^>,G^<$ in eqns. (\ref{kernelkappa},\ref{kernelsigma}), hence they are   included in the general considerations of section (\ref{sec:noneLeff}). To see this, let us consider the (ALP)-gluon interaction vertex $g_s a(x) G^{\mu \nu b}(x)\widetilde{G}_{\mu \nu b}(x)$. The process $a+\mathrm{gluon} \leftrightarrow q \overline{q}$ is contained in the correlation function $\langle G \widetilde{G} G \widetilde{G}\rangle$ at two loops, with one gluon propagator featuring a $q\overline{q}$ self-energy loop, this is the QCD equivalent  of  the second diagram in fig.(\ref{fig:moreloops})  featuring a fermion loop correction to the propagator of the gauge boson. Using Cutkosky's cutting rules it is a simple exercise to see that the rate for the scattering process $a+\mathrm{gluon} \leftrightarrow q \overline{q}$ is given by the imaginary part of the two loop diagram where the cut goes through the $q\overline{q}$ loop. Similarly for the processes $a +\mathrm{gluon} \leftrightarrow 2 \mathrm{gluons}$ which corresponds to a gluon loop for a gluon self-energy.  This is the thermalization rate that enters in the Boltzmann equation in ref.\cite{masso} or the cross section in ref.\cite{turnerther}.

      \item \textbf{Mixed cold and hot components:}  An important corollary of the Langevin-like equation of motion (\ref{langevin}) are the general results (\ref{eneasy},\ref{enerdens}) which entail that the energy density of (ALP) feature a  mixture of  cold and hot components, the  cold component is determined  by oscillatory coherent condensate resulting from misaligned initial conditions  and the hot corresponds to the thermalized part, which  is determined by the ``noise'' term in the Langevin equation, with proportions varying in time  as $ \simeq  (cold)e^{-\Gamma t}+ (hot) (1-e^{-\Gamma t}) $. The damping of the cold component is a consequence of the ``friction'' term in the equations of motion determined by the imaginary part of the self energy, and  the growth rate of the hot component, namely the thermalization rate,     is related to the damping rate of the cold  component by the fluctuation dissipation relation. The cold component originates in the coherent oscillations resulting from a ``misaligned'' initial condition, whereas the hot component results from the approach to thermal equilibration with the bath.

    \item \textbf{Novel exotic phases:?} For (ALP)-photon coupling, the real part of the self-energy reveals two important features: because the coupling $g$ has dimensions $1/(energy)$ the interaction Lagrangian density is non-renormalizable. As a result we find that the effective action must necessarily include higher derivative terms of the form $C \,(\partial_\mu \partial^\mu a(x))^2$ with $C$ a constant that absorbs the logarithmic ultraviolet divergence (\ref{sigcero}). We have (arbitrarily) set $C=0$ to establish contact with the usual Lagrangian proposed for (ALP), but this clearly implies a fine-tuning. Furthermore, the finite temperature part of the self-energy features  the high temperature limit (\ref{hiT}) which yields an effective temperature dependent mass squared given by eqn. (\ref{Tcdef}). The power of temperature $\propto T^4$ is a consequence of the non-renormalizable coupling with mass dimension $-2$.   The negative sign yields the opposite behavior compared to scalar theories with (second order) phase transitions,  the physical origin of the negative sign eludes these authors. We conjecture that the sign is a result of the coupling to a pseudoscalar composite operator with vector fields, but such conjecture awaits confirmation by comparing to other pseudoscalar couplings such as those shown in eqn. (\ref{inter}), which is beyond the original scope of this study.  This effective mass squared \emph{suggests} the possibility of an \emph{inverted} phase transition with $m^2(T) < 0$ for $T> T_c$ signalling an instability towards a phase of lower free energy. Such instability entails that non-linearities in the (ALP) effective Lagrangian are relevant, these may be associated with a potential for the (ALP) field, or from higher orders in the effective action, for example a term of the form $\simeq g^4 a^4$ (with the various branch labels $\pm$), which because of the non-renormalizable nature of the coupling will feature  the largest scale in the loop to the fourth power and may conspire with the quadratic term to stabilize the theory. The emergence  of these non-linearities in higher orders of the effective action merit further study.  The main result of the energy density (\ref{enerdens}) is valid only for $T<T_c$ because the analysis relies on the perturbative renormalization of the frequencies, so that $\Omega_k$ in (\ref{enerdens}) is real.

        Taken together, this instability in combination with higher derivative terms \emph{may} lead to novel exotic inhomogeneous phases for $T>T_c$ of the Lifshitz type\cite{lifshitz}. The possibility of high temperature instabilities and novel phases are worthy of a more detailed and deeper study including other types of pseudoscalar interactions, which is beyond the scope of this article.

        \item \textbf{QED vs. QCD:} Although this discussion has focused on (ALP)-photon coupling, a similar conclusion can be drawn for (ALP)-gluon coupling $g_s a(x) G^{\mu \nu b}(x)\widetilde{G}_{\mu \nu b}(x)$, since $g_s$ also has dimensions of $(energy)^{-1}$.  To lowest order in the strong coupling $\alpha_s$  the correlation function $\langle G \widetilde{G} G \widetilde{G} \rangle$ is a gluon loop  and yields a similar   high temperature dependence of the effective mass squared  $ \propto g^2_s T^4$ on dimensional grounds, and a zero temperature  logarithmic  ultraviolet divergence which requires a higher derivative counterterm. Although similar to the QED case, the actual contributions from gluon loops must be studied in detail because the non-abelian nature may lead to cancellations which these simple arguments may not capture. The study of the QCD contribution from gluons must necessarily focus on temperatures scales above the deconfinement temperature $\simeq 150 \,\mathrm{MeV}$, which requires hard-thermal loop resummations\cite{htl,lebellac} since the light quarks are ultrarelativistic in this temperature range whereas for  $T < 150 \,\mathrm{MeV}$ (ALP) interact with neutral pions. The study of these processes is well beyond the scope of this article but clearly merit further study.

        \item \textbf{Possible cosmological consequences:} While the results obtained above are valid in Minkowski space time, we can \emph{conjecture} on their possible implications in cosmology. The effective squared mass at high temperature (\ref{Tcdef})
             suggests a high temperature inverted phase transition with $m^2(T) < 0$ for $T > T_c$ becoming positive for $T < T_c$, the opposite of the usual behavior in (second order) phase transitions. This in turn  implies that the non-linearities in the ALP (effective) potential are important in the evolution of the coherent condensate, furthermore, the necessity of introducing higher order derivatives to absorb logarithmic ultraviolet divergences  when combined with the high temperature instability may lead to novel inhomogeneous phases, such as Lifshitz phases\cite{lifshitz} with the possible generation of inhomogeneities associated with the dark matter component that are not a consequence of inflationary fluctuations.

            The time evolution of the energy density yielding a mixture of a cold and a hot component (\ref{enerdens}) gives rise to the interesting possibility that the ``warmth'' of this dark matter candidate evolves in time from a colder to a hotter component, the weight of each component is determined by the relaxation rate and the time scale. Hence it is possible that for a specific set of parameters (coupling and mass) the dark matter component is cold at the time of recombination but warms up as time evolves towards a warmer component, thereby yielding (ALP)'s as a warm dark matter candidate in the most recent Universe. This possibility has potentially important consequences for galaxy formation since an (ALP) which is a warm dark matter candidate  may help to solve the core vs. cusp problem in dwarf galaxies.

            Furthermore, if the (ALP) is an ultralight dark matter candidate, it can become an ultrarelativistic component even for a temperature $\simeq T_{cmb} \simeq 0.1  eV$ at the time of recombination, which then contributes to $N_{eff}$ the effective number of relativistic species. As the interaction with the cosmic microwave background continues after recombination until today, the decay of the coherent condensate component and   thermalization may affect the signal on birefringence if it is a consequence of the interaction of the CMB with a pseudoscalar field\cite{komatsu}.

        \item \textbf{Caveats:} In this article we have studied the effective action and its consequence in Minkowski space-time as a prelude towards a more comprehensive study including cosmological expansion which will be addressed in future work. Cosmological expansion introduces several important modifications: in the evolution of the condensates (coherent states) from misaligned initial conditions, dilution of the population and time dependent relaxation rates\cite{herring} among the most obvious ones. In the regime when the cosmological expansion rate $H(t)$ is much smaller than the relaxation rate, we expect an adiabatic treatment (see ref.\cite{herring})  to be reliable. However, in this case we would expect that (ALP) would completely thermalize with the (CMB) after recombination  and would feature the (CMB) temperature \emph{today}. In obtaining the effective action we have traced over the (CMB) degrees of freedom therefore we cannot assess at this stage whether the \emph{back reaction} $a \rightarrow 2\gamma$ would induce distortions in the (CMB) power spectrum. Such distortion would impose severe constraints on the coupling and mass of the (ALP) fields since these determine the relaxation rate. If, on the other hand the relaxation rate is much smaller than $H(t)$ we would expect that the thermal (hot) (ALP) population today would be rather small.
            In our treatment we have assumed the initial (ALP) density matrix to describe a misaligned \emph{vacuum} state, described by a coherent state of a free field vacuum. This initial state neglects any population that could have been produced earlier, such as a produced thermally from QCD processes\cite{turnerther,masso,buch} or even processes beyond the standard model or during inflation. A thermal initial condition can be accounted for, including misalignement,  simply by proposing a coherent state built from a thermal density matrix. Such modification will result in new contributions to the correlation functions and energy density from the initial averages with the Wigner function or alternatively with the initial density matrix. In particular this scenario  would yield another thermal contribution to the energy density originating in the initial density matrix of the (ALP) field, therefore the results obtained in this study provide a lower bound on the (ALP) energy density.

   \end{itemize}

\vspace{1mm}

\section{Conclusions:}\label{sec:conclusions}

We studied the non-equilibrium dynamics of a pseudoscalar (ALP) particle weakly coupled   to ``environmental'' degrees of freedom in thermal equilibrium in Minkowski space-time as a prelude towards extending the methods to cosmology.  We considered  a generic coupling $g a(x) \mathcal{O}(x)$ with $\mathcal{O}$ a pseudoscalar composite operator of the bath degrees of freedom without adopting a particular set of parameters, couplings and (ALP) mass or bounds on them but only assuming a   weak coupling between the (ALP) and the standard model degrees of freedom. Our focus in this article is to obtain the (ALP)   effective action and equations of motion and to explore their consequences for general   couplings and mass.

 By considering the time evolution of an initial density matrix for the (ALP) and environmental fields in the in-in or Schwinger-Keldysh formulation, we obtained the reduced density matrix for the (ALP) by tracing over the environmental fields. The time evolution of the (ALP) reduced density matrix is determined by the non-equilibrium effective action, which we obtain up to $\mathcal{O}(g^2)$ in the weak coupling $g$ but to \emph{all orders} in the couplings of the environmental fields to any other field (different from the (ALP)) within or beyond the standard model.  The effective equations of motion for the (ALP) field obtained from the in-in effective action are causal   Langevin equations with a (non-local) self-energy  and a Gaussian stochastic noise term whose power spectra fulfill the fluctuation-dissipation relation. The initial density matrix for the (ALP) field implements a ``misaligned'' initial condition. The effective Langevin equations of motion show that the processes that lead to the damping of the coherent condensate are the same that lead to thermalization with the environment as a direct result of the fluctuation dissipation relation. Whereas previous studies either focused on the ``friction'' term in the equations of motion of the coherent condensate, or on thermalization via Boltzmann equations, the non-equilibrium effective action and Langevin equation obtained in this study establishes a bridge between both aspects linking them via the fluctuation dissipation relation, a hitherto unrecognized but important aspect of coupling to an environment and shows that both occur on similar time scales. Damping of the coherent misaligned expectation value and thermalization with the environment emerge naturally from the effective Langevin equations of motion, and for generic environments we find that the total energy density features a mixture of a cold and hot components: $\mathcal{E}(t) = (\mathrm{cold})e^{-\Gamma t} + (\mathrm{hot})(1-e^{-\Gamma t})$ the cold component is a consequence of the coherent oscillations from misalignment and the hot component from thermalization with the bath. The relaxation rate $\Gamma$ is determined by the imaginary part of the self-energy. The damping of the cold and the growth of the hot components are a direct consequence of the fluctuation-dissipation relation.

This time dependent energy density \emph{may} provide a compelling dark matter scenario wherein the ``warmth'' of the dark matter evolves in time from colder to hotter. This is one of the important results of our study.

As a specific example we study (ALP)-photon coupling with $\mathcal{O} = \vec{E}\cdot\vec{B}$ where the radiation field represents the (CMB) after recombination when photons can be treated as free and massless (vanishing plasma frequency). This is a non-renormalizable interaction, the one loop contribution to the (ALP) self-energy features ultraviolet divergences  that necessitate higher derivative terms in the effective action, of the form $C (\partial_\mu \partial^\mu a(x))^2$. The long wavelength relaxation rate $\Gamma = \frac{g^2 m^3_a}{64\pi}\Big[1 + 2 n(m_a/2) \Big]$ features a large enhancement for $T\gg m_a$ which is substantial even for the (CMB) temperature $\simeq 10^{-4} eV$ if the (ALP) is a light dark matter candidate with $m_a \lesssim \mu eV$ and even more so if it is an ultralight candidate with $m_a \simeq 10^{-20} eV$.   We find that the high temperature limit of the self-energy yields a  temperature dependent effective mass squared $m^2_a(T) = m^2_a(0)\Big[ 1- (T/T_c)^4 \Big]$ with $T_c \simeq \sqrt{m_a(0)/g}$ suggesting a possible \emph{inverted phase transition} with a negative mass squared for $T>T_c$ which when combined with higher derivative terms in the effective action may lead to the possibility of novel exotic phases.

This study has revealed aspects that have not been previously discussed, such as the necessity of higher derivative operators, the high temperature correction to the mass which suggests a possible inverted phase transition,  and that a misaligned initial condition naturally leads to an energy density that features a mixture of cold and hot components with fractions that depend on time through the relaxation rate, with the cold component diminishing and the hot component increasing in time. If (ALP) are suitable dark matter candidates this mixed cold-hot component may lead to interesting cosmological consequences:  for structure formation the ``warmth''  of the dark matter, a consequence of the cold and hot components, \emph{may} help in solving the core vs cusp problem, furthermore, the hot component may provide  a contribution to the effective number of relativistic degrees of freedom at recombination, and the continued interaction between the (ALP) and the (CMB) post recombination until today may affect a birefringence signature if it is a consequence of a coupling of the (CMB) to a pseudoscalar field. These results may also point to possibly alternative bounds on the couplings and mass of (ALP)s.

 The next step is to extend the methods implemented here  to the realm of an expanding cosmology as well as other possible interactions which will be the  focus of  future work.

\acknowledgements
  The authors gratefully acknowledge  support from the U.S. National Science Foundation through grant   NSF 2111743.

\appendix

\section{Spectral density for $\vec{E}\cdot \vec{B}$ coupling.}\label{app:ebcoup}
We begin with the quantization  of the gauge field within a volume $V$ eventually taken to infinity,
\be \vec{A}(x)= \frac{1}{\sqrt{V}}\sum_{\vk,\lambda=1,2} \frac{\hat{\epsilon}_{\vk,\lambda}}{\sqrt{2k}}\,\Big[d_{\vk,\lambda}\,e^{-ik\cdot x}+d^\dagger_{\vk,\lambda}e^{ik\cdot x}\Big]\,, \label{aquant}\ee  where  $\hat{\epsilon}_{\vk,\lambda}$ are the transverse polarizaton vectors chosen to be real. From  eqns (\ref{Ggfd},\ref{Glfd}) we need the correlation functions
\bea G^>(x-y) & = & \langle \vec{E}(x)\cdot \vec{B}(x)\vec{E}(y)\cdot \vec{B}(y)\rangle \,,\label{Ggeb} \\
G^<(x-y) & = & \langle \vec{E}(y)\cdot \vec{B}(y)\vec{E}(x)\cdot \vec{B}(x)\rangle = G^>(y-x) \,, \label{Gleb} \eea where  we now refer to $\langle (\cdots ) \rangle$ as averages in the thermal density matrix of free field photons.

In the thermal ensemble the expectation value $\langle \vec{E}(x)\cdot \vec{B}(x) \rangle =0$ by parity invariance.      Using Wick's theorem the correlation function
\be \langle \vec{E}(x)\cdot \vec{B}(x)\vec{E}(y)\cdot \vec{B}(y)\rangle = \sum_{i,j}\Big\{ \langle E^i(x)\, E^j(y) \rangle \langle B^i(x)\, B^j(y) \rangle + \langle E^i(x)\, B^j(y) \rangle \langle B^i(x)\, E^j(y) \rangle  \Big\}\,.  \label{correEB}\ee A straightforward calculation yields
\be  \langle E^i(x)\, E^j(y) \rangle  = \langle B^i(x)\, B^j(y) \rangle = \frac{1}{2V}\sum_{\vk} k\,\Big(\delta^{ij}-\hat{\vk}^i \hat{\vk}^j\Big)\,\Big[(1+n(k))\,e^{-ik\cdot(x-y)} + n(k) \, e^{ ik\cdot(x-y)}\Big] \,, \label{eecor}\ee similarly
\be \langle E^i(x)\, B^j(y) \rangle  = - \langle B^i(x)\, E^j(y) \rangle = -\frac{1}{2V} \sum_{\vk} k\, \Big(\hat{\epsilon}^i_{\vk,1}\,\hat{\epsilon}^j_{\vk,2}-\hat{\epsilon}^i_{\vk,2}\,\hat{\epsilon}^j_{\vk,1} \Big)\, \Big[(1+n(k))\,e^{-ik\cdot(x-y)} + n(k) \, e^{ ik\cdot(x-y)}\Big] \,,\label{ebcorr} \ee where $n(k) = 1/(e^{\beta k} -1)$. Combining the two terms in (\ref{correEB}) we find
\bea G^>(x-y) &  = &  \frac{1}{4V^2} \sum_{\vk}\sum_{\vp} kp(1-\hat{\vk}\cdot \hat{\vp})^2 \Bigg\{ \Big[(1+n(k))\,e^{-ik\cdot(x-y)} + n(k) \, e^{ ik\cdot(x-y)}\Big]\nonumber \\ & \times & \Big[(1+n(p))\,e^{-ip\cdot(x-y)} + n(p) \, e^{ ip\cdot(x-y)}\Big]\Bigg\}\,.  \label{Ggfi}\eea
Expanding the product, we perform the following change of variables in the various terms: 1) in the term $n(k)n(p)$: $\vk \rightarrow -\vk,\vp \rightarrow -\vp$, 2) in the term with $(1+n(k))n(p)$: $\vp \rightarrow -\vp$, 3) in the term with $n(k)(1+n(p))$: $\vk \rightarrow -\vk$, yielding in the infinite volume limit
\be G^>(x-y) = \int \frac{dq_0}{2\pi}\int \frac{d^3q}{(2\pi)^3}\,\rho^>(q_0,q)\, e^{-iq_0(t-t')}\,e^{i\vec{q}\cdot(\vx-\vy)}\,,\label{ggro} \ee where
\bea \rho^>(q_0,q) & = &  \frac{\pi}{2}\int \frac{d^3k}{(2\pi)^3}k |\vq-\vk|\Bigg\{\Big(1-\frac{\vk}{k}\cdot\frac{\vq-\vk}{|\vq-\vk|} \Big)^2\,\Big[(1+n(k))(1+n(|\vq-\vk|))  \delta(q_0-k-|\vq-\vk|)\nonumber \\ & + &  n(k)n(|\vq-\vk|)\,\delta(q_0+k+|\vq-\vk|)\Big]\nonumber \\
& + & \Big(1+\frac{\vk}{k}\cdot\frac{\vq-\vk}{|\vq-\vk|} \Big)^2\,\Big[(1+n(k))n(|\vq-\vk|)\delta(q_0-k+|\vq-\vk|) \nonumber \\ & +  & n(k)(1+n(|\vq-\vk|)) \delta(q_0+k-|\vq-\vk|)\Big]\Bigg\}\,.\label{roge} \eea Writing
\be G^<(x-y) = \int \frac{dq_0}{2\pi}\int \frac{d^3q}{(2\pi)^3}\,\rho^<(q_0,q)\, e^{-iq_0(t-t')}\,e^{i\vec{q}\cdot(\vx-\vy)}\,,\label{glro} \ee and using the relation (\ref{Gleb}) we find that $\rho^<(q_0,\vq) = \rho^>(-q_0,-\vq)$, however the sign change in $\vq$ can be compensated by $\vk \rightarrow -\vk$ inside the k-integral with the final result
\be \rho^<(q_0,\vq) = \rho^>(-q_0,\vq) \,, \label{iden}\ee  furthermore, using the identity $(1+n(w)) = e^{\beta w} n(w)$ and using the various delta functions in the definition of $\rho^>$ we find
\be \rho^<(q_0,\vq) = e^{-\beta q_0}\,\rho^>(q_0,\vq)\,, \label{inden2}\ee which is the Kubo-Martin-Schwinger relation, thereby confirming the general results  (\ref{KMS}). The spectral density is given by (see eqn. (\ref{specOs})) $\rho(q_0,q) = \rho^>(q_0,q) - \rho^<(q_0,q)$ with
\bea && \rho(q_0,q)    =     \frac{\pi}{2}\int \frac{d^3k}{(2\pi)^3} \frac{1}{k w} \Bigg\{\big(k w + k^2 -\vk\cdot\vq \big)^2\,[1+n(k)+n(w)]\big(\delta(q_0-k-w)-\delta(q_0+k+w) \big) \nonumber \\ & + & \big(k w - k^2 +\vk\cdot\vq \big)^2\,(n(w)-n(k))\big(\delta(q_0-k+w)-\delta(q_0+k-w) \big)    \Bigg\}~~;~~ w = |\vq-\vk|\,.  \label{rhoeb} \eea
The spectral density is calculated by implementing the following steps:
\be \int \frac{d^3k}{8\pi^3} = \int^\infty_0 k^2 \frac{dk}{4\pi^2} d(cos(\theta)) ~~;~~ w = |\vq-\vk| = \sqrt{q^2+k^2-2kq\cos(\theta)}~~;~~ \frac{d(\cos(\theta))}{w}  = -\frac{d\,w}{kq} \,.\label{steps}\ee Carrying out the integrations, which are facilitated by the delta function constraints we find

\begin{equation}
    \rho(q_0,\vec{q})
    = \frac{(Q^2)^2}{32\pi}\,\Bigg\{\Bigg(1 + \frac{2}{\beta q}\,\ln\Bigg[\frac{1-e^{-\beta \omega^I_+}}{1-e^{-\beta \omega^I_-}} \Bigg]\Bigg)\,\Theta(Q^2)  + \frac{2}{\beta q}\, \ln\Bigg[\frac{1-e^{-\beta \omega^{II}_+}}{1-e^{-\beta \omega^{II}_-}} \Bigg]\,\Theta(-Q^2) \Bigg\}\, \mathrm{sign}(q_0)\,, \label{rhofi}
\end{equation} where
\be Q^2= q^2_0 - q^2 ~~;~~ \omega_\pm^{(I)} = \frac{|q_0| \pm q}{2}~~;~~ {\omega}_\pm^{(II)} = \frac{q \pm |q_0|}{2}\,.\label{Q2omegas}\ee

\section{Finite temperature contribution to $\Sigma_R$}\label{app:finiTsig}

 \begin{equation}
    \Sigma_R^{(T)}(\nu,k)
    = \frac{g^2\,T}{32\pi^2\,k}\, \mathcal{P} \int_{-\infty}^\infty   \frac{(k^2_0-k^2)^2}{\nu - k_0}
      \ln\Big[\frac{1 - e^{-\beta \omega_+}}{1 - e^{-\beta \omega_-}}\Big]\, dk_0
    \equiv  \frac{g^2\,T}{32\pi^2 \, k}\, \mathcal{I}(\nu,k) ~~;~~  \omega_\pm = \Big|\frac{k \pm k_0}{2}\Big|\,.  \label{sigT}
\end{equation}
Since the argument of the logarithm is odd under $k_0 \rightarrow -k_0$, it follows that $\mathcal{I}$ can be written as
\be \mathcal{I}(\nu,k) =\mathcal{P} \int^\infty_0 \frac{2k_0(k^2_0-k^2)^2)}{k^2_0-\nu^2}\,\ln\Bigg[\frac{1-e^{-\frac{\beta}{2}|k_0-k|}}{1-e^{-\frac{\beta}{2} (k_0+k)}} \Bigg]\,dk_0= \mathcal{I}_1+ (\nu^2-k^2)^2\,\mathcal{I}_2 \,,\label{isplit}  \ee where
\bea \mathcal{I}_1 & = &  \mathcal{P} \int^\infty_0  {2k_0(k^2_0-\nu^2+ 2(\nu^2-k^2))} \,\ln\Bigg[\frac{1-e^{-\frac{\beta}{2}|k_0-k|}}{1-e^{-\frac{\beta}{2} (k_0+k)}} \Bigg] \,dk_0 \nonumber \\ \mathcal{I}_2 & = &  \mathcal{P} \int^\infty_0 \frac{2k_0}{k^2_0-\nu^2}\,\ln\Bigg[\frac{1-e^{-\frac{\beta}{2}|k_0-k|}}{1-e^{-\frac{\beta}{2} (k_0+k)}} \Bigg]\,dk_0 \,. \label{iis}  \eea

Using the results
\begin{align}
    \int_0^\infty x^n \ln\Big[1 - e^{-(x+y)}\Big] dx & = - \Gamma(n+1) Li_{2+n}(e^{-y})
    \label{eqn:I1-integral-plus}
    \\
    \int_0^\infty x^n \ln\Big[1 - e^{-|x-y|}\Big] dx & = (-1)^n \Gamma(n+1) Li_{n+2}(e^{-y}) - 2 \sum_{i=0}^{[\frac{n}{2}]} \binom{n}{2i} \Gamma(1 + 2i) \zeta(2 + 2i)\, y^{n-2i}\,,
    \label{eqn:I1-integral-minus}
\end{align}
where $Li$ is the polylogarithm, we find
\begin{equation}
    \mathcal{I}_1 = - \frac{4\pi^2 k}{3 \beta} \Bigg((\nu^2-k^2) + \frac{8 \pi^2}{5 \beta^2}\Bigg)\,.
\end{equation}

For $\mathcal{I}_2$, we first write it as
\begin{equation}
    \mathcal{I}_2 = \mathcal{P} \int_0^\infty dk_0 \Big(\frac{1}{k_0 - \nu} + \frac{1}{k_0 + \nu}\Big) \ln\Bigg[\frac{1 - \exp\Big(- \beta \frac{|k_0 - k|}{2}\Big)}{1 - \exp\Big(- \beta \frac{k + k_0}{2}\Big)}\Bigg]\,.
\end{equation}
Note that
\be  \ln(1-x)   =   \sum_{n=1}^\infty \Big(-\frac{1}{n} x^n \Big)\,,       \ee and
\be \mathcal{P} \int^{\infty}_0 \frac{dx}{x+z}\,\Bigg( - \frac{1}{n}\,e^{-n(x+y)}\Bigg) = \frac{1}{n}\,e^{-n(y-z)}\,Ei(-n z)\ee
\be \mathcal{P} \int^{k}_0 \frac{dx}{x+z}\,\Bigg( - \frac{1}{n}\,e^{-n(k-y)}\Bigg) = -\frac{1}{n}\,e^{-n(y+z)}\,\Big[ -Ei(n z)+ Ei(n(y+z)) \Big]\ee
\be \mathcal{P} \int^{\infty}_k \frac{dx}{x+z}\,\Bigg( - \frac{1}{n}\,e^{-n(x-y)}\Bigg) =  \frac{1}{n}\,e^{ n(k+z)}\, Ei(-n(y+z))\,. \ee The exponential integral function features a useful representation,
\be Ei(x) = \gamma + ln(|x|) + \sum_{n=1}^{\infty} \frac{x^n}{n\,n!}\,, \ee where $\gamma$ is Euler's constant.
This expansion  allows us to extract the low and high temperature limits, yielding the high temperature behavior for $T \gg \Omega_k$
\be {\Sigma^T_R(\Omega_k,k)}  =  {g^2 T^4}   \Bigg[ -\frac{\pi^2}{15} -  \frac{m^2_a}{24\,T^2}+ \frac{m^4_a }{32\,T^4} \, \bigg( 1 - \gamma + \ln\bigg[\frac{4\pi T}{m_a}\bigg] - \frac{\Omega_{\vec{k}}}{k} \ln\bigg[\frac{\Omega_{\vec{k}} + k}{m_a}\bigg] \bigg)
 +\cdots \Bigg]\,.\label{hiTapp} \ee In the low temperature limit $T \ll m_a, k$ we find
 \be\Sigma^T_R(\Omega_k,k) =   g^2 T^4  \,  \Bigg[\frac{4\pi^2}{45} \frac{k^2 }{m^2_a\,}+ \frac{32\pi^4\,m_a}{63}\Big(1 + 4 \frac{k^2}{m^2_a}+ \frac{16}{5} \frac{k^4}{m^4_a}  \Big)\,\frac{T^2}{m^2_a}+ \cdots  \Bigg]\,.\label{loTapp} \ee

\end{document}